\documentclass[showpacs,preprintnumbers,amsmath,amssymb,nofootinbib]{revtex4}

\usepackage{graphicx}
\usepackage{bm}
\usepackage{epstopdf}
\usepackage{float}
\usepackage{booktabs}
\usepackage{dcolumn}
\usepackage{bm}
\usepackage{mathrsfs}
\usepackage{amsmath}
\usepackage{footnote}
\usepackage{accents}
\usepackage{mathtools}
\usepackage{tensor}
\usepackage[titletoc]{appendix}
\usepackage{color}

\begin{document}



\title{Non--singular non--flat universes}
\author{Andrés Felipe Estupiñán Salamanca}
\email{af.estupinan@uniandes.edu.co}
\affiliation{
Departamento de F\'isica,\\ Universidad de los Andes, Cra.1E
No.18A-10, Bogot\'a, Colombia
}
\author{Sergio Bravo Medina}
\email{s.bravo58@uniandes.edu.co}
\affiliation{
Departamento de F\'isica,\\ Universidad de los Andes, Cra.1E
No.18A-10, Bogot\'a, Colombia
}
\author{Marek Nowakowski}
\email{mnowakos@uniandes.edu.co}
\affiliation{
Departamento de F\'isica,\\ Universidad de los Andes, Cra.1E
No.18A-10, Bogot\'a, Colombia
}
\author{Davide Batic}
\email{davide.batic@ku.ac.ae}
\affiliation{
Department of Mathematics, \\
Khalifa University of Science and Technology,
Sas Al Nakhl Campus,
P.O. Box 2533 Abu Dhabi,
United Arab Emirates
}

\date{\today}

\begin{abstract}
 The quest to understand better the nature of the initial cosmological
 singularity is with us since the discovery of the expanding
 universe. Here, we propose several non-flat models, among them the
 standard cosmological scenario with a critical cosmological constant,
the Einstein-Cartan cosmology, the Milne-McCrea universe with quantum
corrections and a non-flat universe with bulk viscosity. Within these
models, we probe into the initial singularity by using different
techniques. Several nonsingular universes emerge, one of the
possibilities being a static non-expanding and stable Einstein universe.

\end{abstract}

\pacs{ }
\maketitle

\section{Introduction}
The nature of the initial cosmological singularity is still not well
understood and the question whether its avoidance depends on a
particular model or is due to quantum correction effects is not
satisfactorily answered. It is therefore of some interest to gain
more insight by studying several different models in which the
singularity
can be avoided due to different reasons inherent in the model.
For a small list of different cosmological points of view and models see
\cite{FaraoniBook1, FaraoniBook2, FaraoniPaper1, FaraoniPaper2,
  FaraoniPaper3, FaraoniPaper4, FaraoniPaper5}.
From another perspective, such an examination will reveal different
singularity-free universes being perpetually produced in a multi-verse by, say,
eternal inflation. At the same time, such an undertaking will give us
the opportunity to use and compare distinct techniques suitable for the study of
the singularity. In this , we focus on non-flat universes with
different underlying assumptions. We start by recalling the example of
the non-flat Loop Quantum Gravity (LQG) cosmology formulated in terms
of critical maximal densities. We proceed to present the non-flat,
nonsingular universe based on the concept of a critical cosmological
constant $\Lambda$. We extend previous calculations for an arbitrary equation of
state (EOS) and show that for the relativistic EOS which evidently is
more probable at the bounce, $\Lambda$ obeys a quadratic equation in
contrast to a cubic one from the dust case. Next, we study the
Einstein-Cartan cosmologies with and without the cosmological
constant. We employ different tools to probe into the nature of the
singularity. A thermodynamical approach is contrasted with calculations
relying on critical densities applied in LQG.  In this model, we also
examine the possibility of a critical cosmological constant.
Similar techniques and approaches are used in examining the non-flat
Milne-McCrea scenario to which we add quantum corrections. In the latter
case, the quantum corrections are responsible for the absence of an
initial singularity. In the Einstein-Cartan framework, it is not 
clear if it is the quantum nature of the torsion source which makes the
model
singularity free. The reason for that is that the source of torsion
is taken to be of the form $S_{\mu \nu}u_{\alpha}$, where $u$ is the
four-velocity and the spin tensor $S_{\mu \nu}$ is derived from a quantity
which has both, the spin and angular momentum (we will come back to
this point at an appropriate place in the text).  Finally, we also
examine the cosmological non-flat model with a viscous energy-momentum
tensor, restricting ourselves to bulk viscosity. Throughout the text, we
use an inequality resulting from the Friedmann equations which proves
very useful for the study of all non-flat models.  This method
resembles a classical potential method. For cosmological models with a
cosmological constant such a method has been discussed elsewhere
starting with \cite{Grav}.
Similar remarks apply also to other methods we use.  The main aim of the
paper is to use different methods for  a variety of new non-flat models  and compare
their effectiveness and strength.

The paper is organized as follows. First, we briefly review the results obtained
in Loop Quantum Cosmology in which nonsingular universes are obtained via the
quantum-corrected Friedmann equation for both flat and non-flat scenarios. Next, we
explore non-flat universes with a non-zero cosmological constant, where we first
introduce the effective potential method to study the evolution of the early universe.
In section IV, we turn to Einstein--Cartan cosmologies, where we introduce a thermodynamical
approach as a different criteria to determine the avoidance of singularity at the early stages of the universe. Here, 
we also explore the effective potential method for these cosmologies. Some critical densities, 
which are natural to certain approaches in the Einstein-Cartan framework, are discussed and interpreted in section V. Section VI is devoted to the study of model universes based on Newtonian dynamics which have the possibility of including quantum corrections. The sign that these corrections attain in the Friedmann equations drastically changes the early universe evolution, and thus, both possibilities are studied. The evolution of the different models arising in this scenario are evaluated using the critical densities method, the effective potential method and the thermodynamical approach. In Section VII, models with bulk viscosity are reviewed. More precisely, their evolution and the singularity presence/absence are assessed through the effective potential method. We summarized our results in the final section.

\section{Nonsingular universes in Loop Quantum Gravity cosmology}
The search for nonsingular universes and the specific conditions which rule their behaviours has lead to different approaches in the context of quantum gravity. One of those is LQG which is a non-perturbative, background independent model that gives certain structure to space in order to solve the short distance problems of the classical theory. It has been shown that for certain types of universes, it also agrees with the semi-classical limit. The cosmological model that follows from this theory, i.e. loop quantum cosmology (LQC), has been studied using modifications of the gravitational Hamiltonian due to quantum geometry and phenomenological predictions in the context of flat and closed universes. We will shortly revisit the analysis that was done for closed and flat universes following \cite{Ashtekar1, Ashtekar2, Ashtekar3}. This approach gives us insights and serves as motivation to our study of the aforementioned types of universes. The quantum description of the small scales requires a different
interpretation as compared to General Relativity where the evolution
of the underlying manifold is studied through dynamical equations for
the scale factor. In this new interpretation, we deal with evolution equations for the
wave function of a universe. In LQG, the evolution
of the universe is studied by a constrain equation on a wave function,
but it can also be interpreted by some effective equations that are
similar to those of the standard cosmology.  This treatment via an effective equation is done by taking a geometrical formulation of quantum
mechanics,
where a special study of the Hamiltonian is performed. We mention that
the effective equations should recover the nonsingular behaviour reflected in the numerical solutions. In the case of flat universes, one must first define appropriate Ashtekar variables $c$ and $p$, which follow the Poisson bracket $\{ c,p \}=\frac{8}{3}\pi G\beta_{BI}$, where $\beta_{BI}$ is the Barbero-Immirzi parameter and $G_N$ is Newton's constant. The Hamiltoniaconstraint is of the form
\begin{equation}
C_{\rm grav}+C_{\rm matt}=-\frac{6}{\beta_{BI}^{2}}c^{2}\sqrt{|p|}+8\pi G_N \frac{p_{\phi}^{2}}{|p|^{3/2}}=0.
\end{equation}
For the gravitational Hamiltonian, this yields 
\begin{equation}
    H_G=-\frac{6}{\beta_{BI}^2}c^2\sqrt{|p|}.
\end{equation}
Adding the contribution of a massless and free scalar scalar field given by 
\begin{equation}
    H_\phi=8\pi G_N \frac{p_{\phi}^2}{|p|^{3/2}},
\end{equation}
the total Hamiltonian constrain can be written as $16\pi G_N(H_G+H_\phi)$. Defining the Hubble parameter as $H=\dot p/(2p)$ and the matter density for the scalar field as $\rho=p_\phi^2/(2|p|^3)$, the usual Friedmann equation follows, namely 
\begin{equation}
    H^2=\frac{8\pi G_N}{3}\rho.
\end{equation}
Now, by promoting the gravitational Hamiltonian as a quantum operator, a new quantum corrected Friedmann equation yields 
\begin{equation}
    H^2=\frac{8\pi G_N}{3}\rho\left(1-\frac{\rho}{\rho_C^{(LQG)}}\right),
\end{equation}
where $\rho_C^{(LQG)}$ is a critical density given by
\begin{equation}
    \rho_C^{(LQG)}=\frac{3}{8\pi G_N\beta_{BI}^2\mu_0^2}.
\end{equation}
The evolution of the scale parameter given by this equation can be seen to be nonsingular, since the Hubble parameter will take a zero value when $\rho=\rho_C^{(LQG)}$. As we can see, not only does it fix the short distance problem of the theory, but at large scales it reproduces the classical theory. In the case of non-flat universes, the quantum-corrected Friedmann equations take the form 
\begin{equation}
    H^2=\frac{\kappa}{3\rho_{crit}}(\rho-\rho_1)(\rho_2-\rho)+O(v^{-3}),
\end{equation}
where $\kappa=8\pi G$, $\rho_{crit}$ is a constant and $\rho_1$, $\rho_2$ are functions
of $v$  which is a parameter related to the scale factor $a$ via the relation $\frac{\dot{a}^{2}}{a^{2}}=\left(\frac{\dot v}{3v} \right)^{2}$. The latter densities can be considered to be critical values when
$\rho=\rho_1$ or $\rho=\rho_2$ as these cases represent the turning points of the scale factor. As it has been shown in \cite{Ashtekar3}, recollapse occurs when $\rho=\rho_1$ while the quantum bounce occurs when $\rho=\rho_2$. In essence, the quantum corrected Friedmann equations in flat and open universes lead to a nonsingular evolution determined by some critical values of the density.

\section{Critical cosmological constant in non-flat universes}
It is well known that the standard cosmology in spatially flat types
of universes leads to an initial singularity or, in other words, to a
zero value of the cosmic scale factor.
In this section, we will show that for a specific cosmological
constant (called critical cosmological constant, or $\Lambda_c$)
it is possible to arrive at a nonsingular universe in the standard
model.
In the literature there exist different approaches to a critical
cosmological constant which have different meanings.
In \cite{RevModPhys.58.689} and \cite{Kardashev1967,Shklovsky1967,Petrosian1967, Petrosian1968,RowanRobinson1968,Petrosian1969,Petrosian1974,Gunn1975,Tinsley1977} the authors refer to coasting
Eddingon-Lamaitre non-singular models such that having determined
the critical cosmological constant $\Lambda_c$ a universe with
$\Lambda=\Lambda_c(1-\epsilon)$ ($\epsilon > 0$ and small) will
experience a behavior asymptotic to Einstein static model in the
infinite past, but is not static for all time $t$.  On the other hand,
e.g., in the very readable paper \cite{Texas} discussing at length the
effect of the cosmological constant in standard cosmologies, the
authors refer among other to the static Einstein model
when they give a relation between $\Lambda$ and $\rho_0$.  

We start from the Friedmann equations with a cosmological
constant
\begin{equation} \label{F1}
\frac{\ddot{a}}{a}=-\frac{4\pi G_{N}}{3}(\rho + 3p) + \frac{\Lambda}{3}
\end{equation}
\begin{equation} \label{eq:4}
H^2 \equiv \left(\frac{\dot{a}}{a}\right)^{2}=\frac{8\pi G_{N}}{3} \rho + \frac{\Lambda}{3}-\frac{k}{a^2R_{0}^{2}},
\end{equation}
from which, after some manipulations, we will arrive at our critical constant model. 
Using the parametric solution for the density, i.e. $\rho=\rho_0a^{-3\gamma}$, the general form of the second Friedmann equation above yields
\begin{equation}
    \dot a ^2 - \frac{\kappa}{3}\rho_0a^{2-3\gamma}-\frac{\Lambda}{3}a^2=-\frac{k}{R_0^2},
    \label{EffE1}
  \end{equation}{}
 where we assumed an equation of state (EOS) of the form $p=(\gamma-1)\rho$ while the scale factor $a(t)=R(t)/R_0$ with $R_0=R(t_0)$ arises from the standard Friedmann-Robertson-Walker metric
 \begin{equation} \label{FRW}
ds^2=dt^2-R^2(t)\left[\frac{dr^2}{1-kr^2} +r^2d\theta^2+r^2\sin^2\theta
d\phi^2\right].
\end{equation}
Defining the scale factor dependent effective potential
\begin{equation}
    V(a)=- \frac{\kappa}{3}\rho_0a^{2-3\gamma}-\frac{\Lambda}{3}a^2,
    \label{pot}
\end{equation}
the Friedmann equation  (\ref{EffE1}) can be written as follows
\begin{equation}
    \dot a^2+ V(a)=-\frac{k}{R_0^2}.
    \label{classP}
\end{equation}
This equation resembles the effective potential approach of the two-body problem in classical mechanics characterized by the equation
\begin{equation}
 \frac{\dot {r}^2}{2} + U_{eff}(r)=E,
\end{equation}
where $E$ denotes the energy per unit mass of the particle. It is important to observe that the dynamics will be allowed in those regions where the motion reality inequality $E-U_{eff}(r) \geq  0$ is satisfied. In other words, this allows a qualitative analysis of the orbits. If we define $\varepsilon=-k/R_0^2$ whose sign depends whether $k=-1,0,1$, the possible values of $a$ are given by the inequality 
\begin{equation}
    \varepsilon-V(a)\geq 0.
  \end{equation}
 In passing we mention that in \cite{Texas} and \cite{Sonego} the
 potential method for models with a cosmological constant has been
 discussed in some depth with a different focus than ours.
 
It is possible to find a critical value for the cosmological constant,
say $\Lambda_c$, such that the corresponding universe
with $\Lambda=\Lambda_c(1-\epsilon)$ is a non-singular coasting universe. If we take $\Lambda>0$ and $\gamma>2/3$, the potential
$V(a)$  will approach $-\infty$ for $a \rightarrow 0$ as well as for
$a \rightarrow \infty$. It will also exhibit a local maximum at $a_{max}$ whenever
\begin{equation}
   V^{\prime}(a)= -\frac{8\pi G_N}{3}\rho_0(2-3\gamma)a^{1-3\gamma}-\frac{2\Lambda}{3}a=0.
    \nonumber
\end{equation}
The solution of the above equation reads
\begin{equation} \label{amax}
    a_{max}=\left[\frac{4\pi G_N \rho_0}{\Lambda}(3\gamma-2)\right]^\frac{1}{3\gamma}.
\end{equation}
At this particular value of $a$, the potential can be calculated to be
\begin{equation}
    V(a_{max})= -\frac{\Lambda\gamma a_{max}^2}{3\gamma-2}.
\end{equation}
Taking into account the shape of the potential (\ref{pot}),
nonsingular universes will arise for $k=1$ provided that
\begin{equation}
    V(a_{max}) \geq -\frac{1}{R_0^2}.
    \label{Inqc}
  \end{equation}
It is useful to define the quantity $R_{oc}$ through the relation  $V(a_{max})=-1/R_{oc}$. The emerging universe  is nonsingular in the sense that if the
inequality is satisfied, the scale parameter will have a positive
minimum value which then can extend up to space-like
infinity. $R_{oc}$ can
be readily calculated to be
\begin{equation} \label{R0}
  R_{oc}=\frac{1}{a_{max}}
  \left(\frac{3\gamma-2}{\Lambda\gamma}\right)^{1/2}.
\end{equation}
Note that $R_{oc}$ must then be the maximum value $R_0$ can take so that
(\ref{Inqc}) holds. If $R_0=R_{oc}$, there is only one possible value for the scale parameter, which is interpreted as a static universe. 
\begin{figure}{}
    \centering
    \includegraphics[scale=0.5]{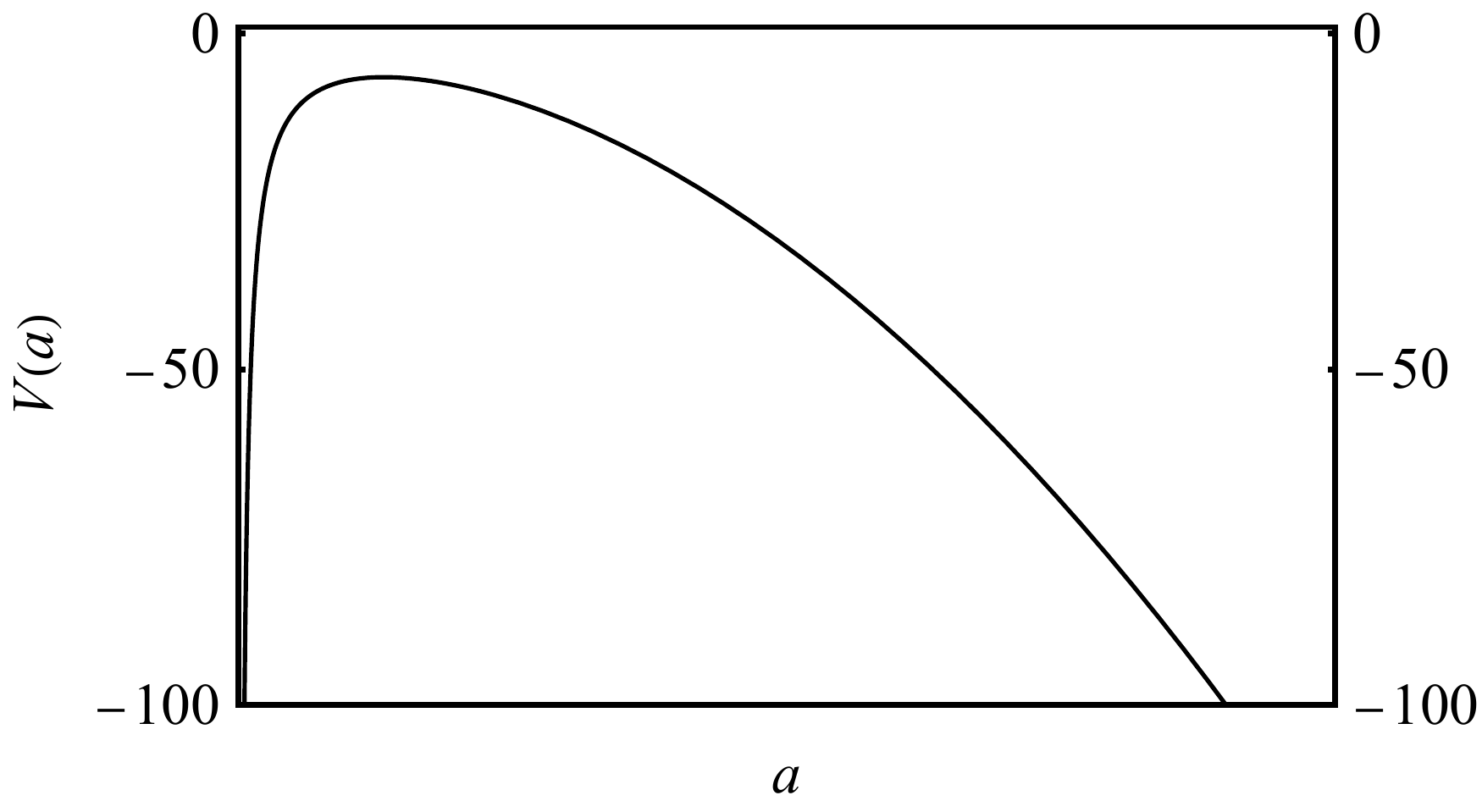} 
    \caption{The plot represents (\ref{pot}) versus $a$, in arbitrary units for $\gamma=1$. nonsingular universes will rise for $\epsilon<V(a_{max})$, where $V(a_{max})$ is represented in the figure as the maximum value of $V(a)$}.
    \label{fig convergencia a 1}
\end{figure}
It is possible to reinterpret this as a condition for $\Lambda$.
At this point the treatments and interpretations of the results in the literature
differ. 
Reference \cite{RevModPhys.58.689} eliminates $R_{oc}$ whereas
\cite{Texas} puts $R_{oc}=1$ defining hereby a fundamental length
scale.  In \cite{RevModPhys.58.689} the authors introduce also a
Hubble parameter $H_0 \neq 0$ which means that the treatment does not
apply strictly to a static model, but to a coasting universe whose
development at later times is non-static.  Reference \cite{Texas} uses
(\ref{R0}) to solve it for $\Lambda$. We first follow the approach of
\cite{RevModPhys.58.689}, but use a general equation of state.

In
order to find the value of $\Lambda_c$ such that non-singular universes
are obtained, we use the second Friedmann equation at $t=t_0$ that is
\begin{equation} \label{eq:6a}
\Omega_{m} + \Omega_{\Lambda}
+\Omega_{k}=1 
\end{equation}
with $k={\rm sgn}(\Omega_{m} + \Omega_{\Lambda}-1)$ and
\begin{equation} \label{eq:7}
\Omega_{m}=\frac{\rho_0}{\rho_{\rm crit}},\quad
\Omega_k= -\frac{k}{R_0^2H_0^2},\quad
\Omega_{\Lambda }=\frac{\rho_{\rm vac}}
{\rho_{\rm crit}},\quad
\rho_{\rm crit}=\frac{3H^2_0}{8\pi G_N},\quad
\Lambda=8\pi G_N  \rho_{\rm vac}.
\end{equation}
This allows to write $R_{oc}$ in the form
\begin{equation} \label{Rcrit2}
    R_{oc}^2=\frac{1}{H_0^2(\Omega_{m}+\Omega_{\Lambda_c}-1)}.
\end{equation}
Combining (\ref{amax}), (\ref{R0}) and (\ref{Rcrit2}) we arrive at
\begin{equation}
    \Omega_{m}+\Omega_{\Lambda_c}-1=\frac{3}{8\pi G_N\rho_{crit}}\left(\frac{\Lambda_c\gamma}{3\gamma -2}\right){\left[\frac{4\pi G_N\rho_0}{\Lambda_c}(3\gamma-2)\right]}^{2/3\gamma}.
\end{equation}
If we make use of the definitions of $\Omega_{m}$ and $\Omega_\Lambda$ as introduced in (\ref{eq:7}),  we finally obtain an equation for $\Omega_{\Lambda_c}$ 
\begin{equation}
    \Omega_{m}+\Omega_{\Lambda_c}-1=3\gamma(3\gamma-2)^{2/3\gamma-1}{\left(\frac{\Omega_{m}}{2}\right)}^{2/3\gamma}\Omega_{\Lambda_c}^{1-2/3\gamma}.
    \label{CritG}
\end{equation}
This equation generalizes the result found in \cite{RevModPhys.58.689}
for any $\gamma$. If we take the special case $\gamma=1$, (\ref{CritG}) becomes
\begin{equation}\label{passo}
    \Omega_{m}+\Omega_{\Lambda_c}-1=3{\left(\frac{\Omega_{m}}{2}\right)}^{2/3}\Omega_{\Lambda_c}^{1/3}
\end{equation}
and employing again the definition of $\Omega_\Lambda$ to (\ref{passo}) yields
\begin{equation}\label{cubozz}
    \frac{\Lambda_c}{12H_0^2\Omega_{m}}-\frac{3}{4}\left(\frac{\Lambda_c}{12H_0^2\Omega_{m}}\right)^{1/3}+\frac{1}{4}\left(\frac{\Omega_{m}-1}{\Omega_{m}}\right)=0.
\end{equation}
It is gratifying to observe that (\ref{cubozz}) can be turned to the same cubic equation whose roots have been discussed extensively in \cite{RevModPhys.58.689}. We conclude this section with the treatment of the case $\gamma=4/3$ which corresponds to the EOS $p=\rho/3$ describing a radiation and relativistic matter dominated universe. The reason of this choice for $\gamma$ resides in the fact that it is more appropriate to model the early stages of the matter in the universe. For this value of $\gamma$, equation (\ref{CritG}) reduces after some trivial manipulations to the quadratic equation
\begin{equation}
    \frac{4\Omega_{\Lambda_c}}{\Omega_{m}}-(\Omega_{m}+\Omega_{\Lambda_c}-1)^2=0.
    \label{M1F}
\end{equation}
The solution to (\ref{M1F}) is then 
\begin{equation}
    \Omega_{\Lambda_c}=\pm2(\Omega_{m})^{1/2}+\Omega_{m}+1.
    \label{CCC}
\end{equation}{}
As a result, we have shown that for a specific cosmological constant
(\ref{CCC}) the spatially non-flat universe will have no initial
singularity. As a side remark, we notice that gravity with a
cosmological constant can violate some assumptions underlying the
global singularity theorems.

Let us now come back to a fully static universe and establish a relation
of $\Lambda$ with the other parameters (notice that we have now
$H=0$).  Reference \cite{Texas}
makes use of equation (\ref{R0}) with $R_{oc}=1$. We supplement their
result by showing that $a_{max}=1$. To this end,  it is useful to use
the second Friedmann equation with $H_0=0$.  This gives explicitly
\begin{equation} \label{extra1}
\rho_0 + \rho_{\rm vac} -\frac{3k}{R_0}\frac{1}{8\pi G_N} =0
\end{equation}
Inserting therein $\rho_0$ and $R_{oc}$ from (\ref{amax}) and
(\ref{R0}) we arrive at
\begin{equation} \label{amax2}
  2(a_{max}^{3\gamma}-1) +3\gamma(1-a_{max}^2)=0
\end{equation}
It is instructive to examine these equation for the two most important
choices, $\gamma=1$ and $\gamma=4/3$. We obtain, respectively
\begin{equation}
  2a_{max}^3-3a_{max}^2 +1=0
\end{equation}
and
\begin{equation}
  2a_{max}^4 -4a_{max}^2+2=0
\end{equation}
One can check that the unique solution for a positive value is
\begin{equation} \label{amax3}
  a_{max}=1
\end{equation}
In other words, if we demand that $a_{max}$ is independent of $\gamma$
we arrive at (\ref{amax3}) also from (\ref{amax2}). Physically, this
result reflects the fact that in a static universe $a=a_{max}$ must be
always one. We can then use (\ref{amax}) with (\ref{amax3}) which
gives
\begin{equation}
  \Lambda=4\pi G_N \rho_0(3\gamma-2)
\end{equation}
or, alternatively (\ref{R0}) and (\ref{amax3}) resulting in
\begin{equation}
  \Lambda=\frac{3\gamma-2}{\gamma R^2_{oc}}
\end{equation}
which can be interpreted as a critical $\Lambda$ with reference to
the static universe.

\section{Einstein-Cartan Cosmologies}
In order to define a cosmological model, one must establish the
interaction between the geometry and matter.
Though this is mostly done through general relativity, there exist other
models that go beyond Einstein gravity. In this section, we choose as
an example the Einstein-Cartan theory of gravity. The motivation
behind such an extension is to introduce spin degrees of freedom into
gravity making an indirect connection with quantum mechanics.
The model provides the simplest mechanism generating a nonsingular
bounce \cite{Poplawski:2018ypb}. A complete description on this model
can be found in
\cite{Medina:2018rnl,article33,osti_4528877}. Here, we will
only introduce the principal components and steps of the theory
necessary to understand
its cosmological implications.

\subsection{Notation}
The most important feature in Einstein-Cartan is the antisymmetric part of the affine connection related to torsion. Torsion can be defined as the antisymmetric part of the connection that is
\begin{equation}
    S_{\mu\nu}^{\;\;\;\alpha}=\Gamma^{\alpha}_{[\mu\nu]}=\frac{1}{2}(\Gamma^{\alpha}_{\mu\nu}-\Gamma^{\alpha}_{\nu\mu}),
\end{equation}
and thus $S_{\mu\nu}^{\alpha}$ is a proper tensor. The connection in
Einstein-Cartan can be written in terms of a contortion tensor $K$ 
\begin{equation}
    \Gamma^{\alpha}_{\mu\nu}=\accentset{\circ}{\Gamma}^{\alpha}_{\mu\nu}-K_{\mu\nu}^{\;\;\;\alpha},
\end{equation}
where the contortion is related to the torsion by
\begin{equation}
    S_{\mu\nu}^{\;\;\;\alpha}=-K_{[\mu\nu]}^{\;\;\;\alpha}=-\frac{1}{2}(K_{\mu\nu}^{\;\;\;\alpha}-K_{\nu\mu}^{\;\;\;\alpha}).
\end{equation}
As done in \cite{Medina:2018rnl}, we will denote the terms with an
over-circle to refer to torsionless objects as in GR. Due to the additional term
in the connection, most of the identities used in GR must be modified. For example, the new expression for the Ricci tensor will be 
\begin{equation}
    R_{\mu\nu}=\accentset{\circ}{R}_{\mu\nu}-\nabla_\lambda K_{\mu\nu}^{\;\;\;\lambda}+\nabla_\mu K_{\lambda\nu}^{\;\;\;\lambda}+K_{\rho\mu}^{\;\;\;\lambda}K_{\lambda\nu}^{\;\;\;\rho}-K_{\mu\nu}^{\;\;\;\lambda}K_{\rho\lambda}^{\;\;\;\rho}.
\end{equation}
One important difference with the theory developed in GR is that this Ricci tensor is not symmetric unlike $\accentset{\circ}{R}_{\mu\nu}$. Hence, we can express the antisymmetric part of the tensor as
\begin{equation}
    R_{[\mu\nu]}=\accentset{\ast}{\mathbf{\nabla}}_\lambda T_{\mu\nu}^{\;\;\;\lambda},
\end{equation}
where the following definitions have been made for the modified tension tensor $T_{\mu\nu}^{\;\;\;\lambda}$, and the star derivative $\accentset{\ast}{\mathbf{\nabla}}_\lambda$ 
\begin{equation}
    T_{\mu\nu}^{\;\;\;\lambda}=S_{\mu\nu}^{\;\;\;\alpha}+S_{\mu\nu}^{\;\;\;\alpha}-S_{\mu\nu}^{\;\;\;\alpha},\quad
   \accentset{\ast}{\mathbf{\nabla}}_\lambda=\nabla_\lambda+2S_{\lambda\alpha}^{\;\;\;\alpha}. 
 \end{equation}
 Upon contraction with the metric tensor, the new Ricci scalar
 multiplied with $\sqrt{-g}$ can be expressed as 
 a sum of three terms: the first term $\sqrt{-g}\accentset{\circ}{R}$ is the usual one already present in GR, the second one contains bilinear terms in $K^2$, i.e. it involves the 
 contortion  while the third term is made up of total derivatives. This makes it clear why the torsion is not a 
 dynamical object and it will not propagate.

\subsection{Equations of motion and field equations}
To derive the equations of motion in the EC framework it is, of
course,  necessary to vary the action as classically done in any
Lagrangian theory. To do so, one must first make a choice of the
Lagrangian which will lead to different equations of motion. This
choice can be made by taking a linear superposition of invariants from
the theory,
but as a standard approach, we begin with the Einstein-Hilbert Lagrangian together
with a matter Lagrangian. The reason behind this choice, apart from
being the most obvious, is due to the fact that it will lead
eventually to a canonical expression of the field equations. We consider the action 
\begin{equation}\label{azione}
    S=\frac{1}{2\kappa}\int d^4x\sqrt{-g}(R+L_m),
\end{equation}
where $L_m$ is a matter Lagrangian, and $R$ is the gravitational Lagrangian given by $R= g^{\mu\nu}R_{\mu\nu}$. A variation of the gravitational part with respect to the inverse metric yields 
\begin{equation}
    \delta_g R= R_{\mu\nu}\delta_g\bold{g}^{\mu\nu}+\bold{g}^{\mu\nu}\delta_gR_{\mu\nu},\quad
    \bold{g}^{\mu\nu}=\sqrt{-g}g^{\mu\nu}
\end{equation}
In this approach, the variation of the Ricci tensor can be found from the contracted
Riemann tensor and making use of the covariant derivative. Splitting off total divergences, one obtains 
\begin{equation}
    \bold{g}^{\mu\nu}\delta_gR_{\mu\nu}=2T_{\rho}^{\nu\lambda}\delta_g\Gamma^{\rho}_{\lambda\nu}.
\end{equation}
The variation of the connection with respect to the metric can be found to be 
\begin{equation}\label{littlehelp}
    \delta_g\Gamma^{\rho}_{\lambda\nu}=\frac{1}{2}g^{\rho\gamma}\left(\nabla_\lambda \delta g_{\nu\gamma}+\nabla_\nu \delta g_{\lambda\gamma}-\nabla_\gamma \delta g_{\lambda\nu}\right).
\end{equation}
Substituting (\ref{littlehelp}) into the geometric part of the integral (\ref{azione}) followed by integration by parts yields  
\begin{equation}
    \int d^4x\sqrt{-g}T_{\rho}^{\nu\lambda}\delta_g\Gamma^{\rho}_{\lambda\nu}=\frac{1}{2}\int d^4x\sqrt{-g}\accentset{\ast}{\mathbf{\nabla}}_\lambda(T_{\nu\mu}^{\;\;\;\lambda}+T_{\nu\;\;\mu}^{\;\;\lambda\;\;}-T_{\;\;\nu\mu}^{\lambda}).
\end{equation}
We note that the part proportional to the variation of the metric is
\begin{equation}
    R_{\mu\nu}\delta_g \bold{g}^{\mu\nu}=\sqrt{-g}\left(R_{\mu\nu}-\frac{1}{2}g_{\mu\nu}R\right)\delta g^{\mu\nu}=\sqrt{-g}G_{(\mu\nu)}\delta g^{\mu\nu},
\end{equation}
where $G_{\mu\nu}$ is the new Einstein tensor defined in the very same
way as in GR but taking into account the differences that the non
symmetrical part of the connection brings. Therefore, taking into
account the other terms of the variation, we arrive at
\begin{equation}
    \frac{1}{\sqrt{-g}}\frac{\delta R}{\delta g^{\mu\nu}}=G_{(\mu\nu)}+\accentset{\ast}{\mathbf{\nabla}}_\lambda(T_{\nu\;\;\;\mu}^{\;\;\lambda}-T_{\;\;\nu\mu}^{\lambda}).
\end{equation}
Repeating the same analysis for the variation with respect to contortion leads to
\begin{equation}
    \frac{1}{\sqrt{-g}}\frac{\delta R}{\delta K_{\mu\nu}^{\;\;\lambda}}=-2T_{\lambda}^{\;\;\nu\mu}.
\end{equation}
Finally, defining the metric energy-momentum tensor $\sigma_{\mu\nu}$
and the source $\tau_{\alpha}^{\;\;\nu\mu}$ due to the matter Lagrangian according to
\begin{equation}
    \sigma_{\mu\nu}\coloneqq -\frac{2}{\sqrt{-g}}\frac{\delta L_m}{\delta g^{\mu\nu}},\quad
    \tau_{\alpha}^{\;\;\nu\mu}\coloneqq \frac{1}{\sqrt{-g}}\frac{\delta L_m}{\delta K_{\mu\nu}^{\;\;\;\alpha}},
\end{equation}
allows to cast the field equations into the form
\begin{eqnarray}
    G_{\mu\nu}+\accentset{\ast}{\mathbf{\nabla}}_\lambda(T_{\nu\;\;\;\mu}^{\;\;\lambda}-T_{\nu\mu}^{\;\;\;\lambda}-T_{\;\;\nu\mu}^{\lambda})&=&\kappa\sigma_{\mu\nu},\\
    S_{\lambda\nu}^{\;\;\;\mu}+\delta^{\mu}_{\lambda}S_{\nu\alpha}^{\;\;\;\alpha}-\delta^{\mu}_{\nu}S_{\lambda\alpha}^{\;\;\;\alpha}&=&\kappa\tau_{\lambda\nu}^{\;\;\;\mu}.
  \end{eqnarray}
 Let $\Sigma^{\mu\nu}$ be the non-symmetrical total energy-momentum tensor which is defined by 
\begin{equation}
    \Sigma^{\mu\nu}\coloneqq\sigma_{\mu\nu}+\accentset{\ast}{\mathbf{\nabla}}_\lambda(\tau^{\mu\nu\lambda}-\tau^{\nu\lambda\mu}+\tau^{\lambda\mu\nu}).
\end{equation}
With this definition the first field equation becomes 
\begin{equation}
    G_{\mu\nu}=\kappa\Sigma_{\mu\nu}.
\end{equation}
This is the canonical form of the Einstein-Cartan field equations. It is not difficult to see that the following equivalent representation is also possible
\begin{equation}
    \accentset{\circ}{G}_{\mu\nu}=\kappa\bar\sigma_{\mu\nu},
\end{equation}
where 
\begin{equation}
    \bar\sigma_{\mu\nu} \coloneqq \sigma_{\mu\nu}+\kappa\left(-4\tau_{\mu\lambda}^{\;\;\;[\alpha}\tau_{\nu\alpha}^{\;\;\;\lambda]}-2\tau_{\mu\lambda\alpha}\tau_{\nu}^{\;\lambda\alpha}+\tau_{\alpha\lambda\mu}\tau^{\alpha\lambda}_{\;\;\;\;\nu}+\frac{1}{2}g_{\mu\nu}(4\tau_{\lambda\;\;\;[\alpha}^{\;\beta}\tau^{\lambda\nu}_{\;\;\;\;\beta]}+\tau^{\alpha\lambda\beta}\tau_{\alpha\lambda\beta})\right).
\end{equation}
Last but not least, by identifying the canonical energy-momentum tensor as the
physical source and choosing the source of torsion as the spin tensor
multiplied by its velocity it is possible to deduce that
\begin{equation}
    \bar\sigma_{\mu\nu}=\left(\rho+p-\frac{1}{2}\kappa s^2\right)u_\mu u_\nu-\left(p-\frac{1}{4}\kappa s^2\right)g_{\mu\nu}-2(u_{\lambda}+\delta^{\alpha}_\lambda)\accentset{\circ}{\nabla}_\alpha(\langle s^\lambda_{(\mu}\rangle u_{\nu)})
\end{equation}
where $\rho$ and $p$ denote the matter density and pressure, respectively. As a consequence, the field equations take the final form 
\begin{equation}
    \accentset{\circ}{G}_{\mu\nu}=\kappa\bar\sigma_{\mu\nu}.
    \label{EFEEC}
\end{equation}

\subsection{Friedmann equations in EC gravity}
Having derived the Einstein field equations in the presence of a nonsymmetric connection, we move to the derivation of the
Friedmann equations in the framework of the Friedmann–Lema\v{i}tre–Robertson–Walker metric (FLRW) metric introduced in (\ref{FRW}). We first consider the case $k=0$ which will be later generalized to non-flat universes. As done before, we will infer the Friedmann equations from
the time and space components of the generalized Einstein tensor. To do this, we write (\ref{EFEEC}) in an explicit way as follows
\begin{equation}
    \accentset{\circ}{R}_{\mu\nu}-\frac{1}{2}\accentset{\circ}{R}=\kappa\left[\left(\rho+p-\frac{1}{2}\kappa s^2\right)u_\mu u_\nu-\left(p-\frac{1}{4}\kappa s^2\right)g_{\mu\nu}-2\left(u_{\lambda}+\delta^{\alpha}_\lambda\right)\accentset{\circ}{\nabla}_\alpha (\langle s^\lambda_{(\mu}\rangle u_{\nu)}) \right].
\end{equation}
The time component of this equation is 
\begin{equation}
  \accentset{\circ}{R}_{00}-\frac{1}{2}\accentset{\circ}{R}g_{00}=\kappa\left[\left(\rho+p-\frac{1}{2}\kappa s^2\right)u_0 u_0-\left(p-\frac{1}{4}\kappa s^2\right)g_{00}-2(u_{\lambda}+\delta^{\alpha}_\lambda)\accentset{\circ}{\nabla}_\alpha(\langle s^\lambda_{(0}\rangle u_{0)})\right].
\end{equation}
Taking into account the Frenkel condition $s_{\mu \nu}u^{\mu}=0$ together with 
the fact that we are working in comoving coordinates allows to reduce the above equation to the form 
\begin{equation}
    \accentset{\circ}{R}_{00}-\frac{1}{2}\accentset{\circ}{R} g_{00}=\kappa\left[\left(\rho+p-\frac{1}{2}\kappa s^2\right)-\left(p-\frac{1}{4}\kappa s^2\right)\right].
\end{equation}
If we compute the Ricci tensor in the case of the FLRW metric, the first Friedmann equation comes out to be
\begin{equation}\label{Friedmann1-EC}
    H^2=\frac{\kappa}{3}\left(\rho-\frac{1}{4}\kappa s^2\right).
\end{equation}
Repeating the same analysis for the space components leads to 
\begin{equation}
    \accentset{\circ}{R}_{11}-\frac{1}{2}\accentset{\circ}{R} g_{11}=\kappa\left[\left(\rho+p-\frac{1}{2}\kappa s^2\right)u_1 u_1-\left(p-\frac{1}{4}\kappa s^2\right)g_{11}\right],
\end{equation}
from which the second Friedmann equation readily follows
\begin{equation}
    2\frac{\ddot a}{a}+\frac{\dot a^2}{a^2}=
    -\kappa\left(p-\frac{1}{4}\kappa s^2\right).
\end{equation}
The latter can also be written with the help of the first Friedmann equation as
\begin{equation}\label{Friedmann2-EC}
    \frac{\ddot a}{a}=-\frac{\kappa}{6}(\rho+3p-\kappa s^2).
 \end{equation}
It is customary to parametrize the spin density of a fluid of fermions
  with no polarization by the number density of fermions $n_f$ as follows 
\begin{equation}\label{Spin-EC}
    s^2=\frac{1}{8}(\hbar c n_f)^2.
\end{equation}
The Friedmann equations take then the form
\begin{equation}
    H^2=\frac{\kappa}{3}\left[\rho-\frac{1}{32}\kappa(\hbar c n_f)^2\right],\quad
    2\frac{\ddot a}{a}+\frac{\dot a^2}{a^2}=
    -\kappa\left[p-\frac{1}{32}\kappa(\hbar c n_f)^2\right].
\end{equation}
If we define an effective energy density and pressure of the form 
\begin{equation}
    \bar\rho=\rho-\alpha n_f^2,\quad
    \bar p= p- \alpha n_f^2,\quad
    \alpha=\frac{1}{32}\kappa\hbar^2c^2
\end{equation}
the Friedmann equations can be simplified to 
\begin{equation}\label{Fr}
    H^2=\frac{\kappa}{3}\bar\rho,\quad
    2\frac{\ddot a}{a}+\frac{\dot a^2}{a^2}=-\kappa\bar p.
\end{equation}
Note that (\ref{Fr}) can be easily generalized to the case with non-zero curvature. The result reads
\begin{eqnarray}
    \Dot a^2+k&=&\frac{1}{3}\kappa\bar\rho a^2,
    \label{F1}\\
    \Dot a^2+2a\ddot a+k&=& -\kappa \bar pa^2.
    \label{F2}
\end{eqnarray}
Such a form of the Friedmann equations provided by \cite{Poplawski:2018ypb} contains as a special case the classical standard cosmology  in the limit of vanishing spin density.

\subsection{A thermodynamical approach}
\begin{figure}[H]
    \centering
    \includegraphics[scale=0.5]{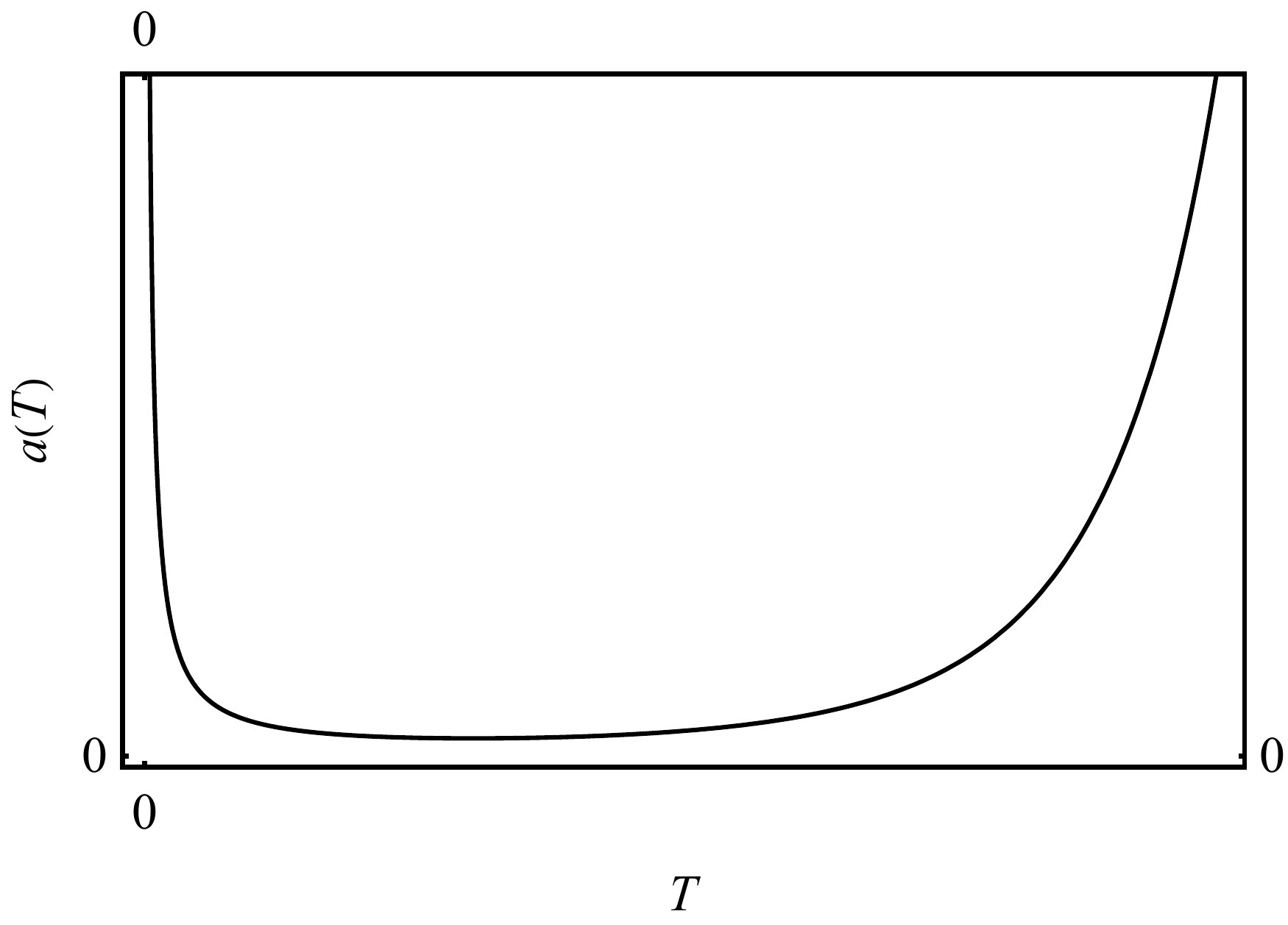} 
    \caption{The plot represents (\ref{aaaaa}) versus $T$ in arbitrary units. $a(T)$ will have a minimum value which represents a bounce.}
    \label{ECT}
  \end{figure}

To get an insight into the consequences of the EC Friedmann equations, 
we can formulate the parametric solution to these equations in
dependence of temperature as done in \cite{cubero2019analysis}. We
begin by multiplying (\ref{F1}) by the scale factor followed by differentiation with
respect to time. We obtain 
\begin{equation}
    \Dot a^3+2a\dot a \ddot a+k\dot a=\kappa\bar\rho a^2\dot a+ \frac{1}{3}a^3\kappa\dot{\bar\rho}.
    \label{4}
\end{equation}
Multiplying (\ref{F2}) by $\dot a$ and subtracting to it (\ref{4}) yields 
\begin{equation}
    3\bar\rho a^2\dot a + 3\bar pa^2\dot a+a^3\dot\rho=0
\end{equation}
which reduces to 
\begin{equation}
    \frac{d}{dt}(\bar\rho a^3)+\bar p\frac{da^3}{dt}=0.
\end{equation}
We can see that this equation shares a close resemblance to that of
the first law of thermodynamics for an adiabatic universe because $a^3$
is proportional to the volume, and $\rho a^3$ could be treated as the
internal energy of the matter and radiation. Using the corrections to
the energy density and pressure, the cosmological law of
thermodynamics takes the form
\begin{equation}
    \frac{d}{dt}\left[(\rho-\alpha n_f^2) a^3\right]+(p- \alpha n_f^2)\frac{da^3}{dt}=0
    \label{YYYYYYY}
\end{equation}
which can be shown to be equivalent to
\begin{equation}
    a^3\dot\rho- 2\alpha a^3n_f\dot n_f +(\rho+p)\frac{da^3}{dt}=0.
    \label{12}
\end{equation}
Since our focus is on the early universe, we can assume that we are in an ultrarelativistic matter regime in kinetic equilibrium. This translates to the EOS $p=\rho/3$ and to the following relations linking the temperature of the universe $T$ to the density and the Fermionic density number, that is $\rho=h_eT^4$ and $n_f=h_{n_f}T^3$ where $h_e= (\pi^2/30)((7/8)g_f+g_b)(k_B^4/(\hbar c)^3)$ and $h_{n_f}=(\zeta(3)/\pi^2)(3/4)g_fk_B^3/(\hbar c)^3)$ (we refer to \cite{cubero2019analysis} for more details). If we implement this information into (\ref{F1}), we can express the first Friedmann equation in the presence of spin and torsion as
\begin{equation}
    \frac{\Dot a^2}{c^2}+k=\frac{1}{3}\kappa (h_eT^4-\alpha h_{n_f}^2T^6) a^2
    \label{13}
\end{equation}
or in the following equivalent form also known as the cosmological law of thermodynamics
\begin{equation}\label{CLT}
h_eT^3dT- \frac{3}{2}\alpha h_{n_f}^2T^5dT +h_eT^4\frac{da}{a}=0.
\end{equation}
If we further divide (\ref{CLT}) by $h_eT^4$ and introduce the new constant $T_{cr}=(2h_e/3\alpha h_{nf}^2)^{1/2}$, we finally arrive at the expression 
\begin{equation}
    \left(\frac{1}{T}- \frac{T}{T_{cr}^{2}}\right)dT +\frac{da}{a}=0.
\end{equation}
Solving this differential equation gives
\begin{equation}
    a(T)=\frac{K}{T}\mbox{exp}\left(\frac{T^2}{2T_{cr}}\right),
    \label{aaaaa}
\end{equation}
where $K$ is a positive but otherwise arbitrary integration constant. From the functional dependence of (\ref{aaaaa}) we immediately conclude that the emerging universe for any temperature $T$ is not singular. Due to its shape and
behaviour it is known as ``Big bounce'' model. Plotting the scale factor (see Figure~\ref{ECT}) reveals that the function
exhibits a local minimum dividing the function into two branches. Clearly, only the decreasing branch is physical as a decreasing $a$
must go with increasing $T$. 
It is interesting to consider as well an application of the thermodynamical method to the
 cosmology resulting from the classic Einstein field equations with a cosmological constant. This simple exercise shows that such a 
method is not always a suitable tool to probe into the nature of the
initial singularity. Without further ado, we recall that Friedmann equations in the presence of a  cosmological constant are
\begin{eqnarray}
    \frac{\dot a^2+k}{a^2}&=&\frac{\kappa\rho+\Lambda }{3},
    \label{FWL1}\\
    \frac{\ddot a}{a}&=&-\frac{\kappa}{6}(\rho+3p)+\frac{\Lambda }{3}.
    \label{FWL2}
\end{eqnarray}
If we consider the combination $a^2$(\ref{FWL1})+$2$(\ref{FWL2}), we obtain an equation which resembles (\ref{F2}), namely 
\begin{equation}
    \Dot a^2+2a\ddot a+k= -\kappa pa^2+\Lambda a^2.
    \label{19}
\end{equation}{}
Moreover, differentiation of $a^3$(\ref{FWL1}) with respect to time leads to
\begin{equation}
    \Dot a^3+2a\dot a \ddot a+k\dot a=\kappa\rho a^2\dot a+ \frac{1}{3}a^3\kappa\dot\rho
    +\Lambda a^2\dot a.
    \label{20}
\end{equation}
Finally, if we multiply (\ref{19}) by $\dot a$ and subtract to it (\ref{20}), we end up with 
\begin{equation}
    \frac{d}{dt}(\rho a^3)+p\frac{da^3}{dt}=0
\end{equation}
and the corresponding cosmological law of thermodynamics takes the form
\begin{equation}
    \dot\rho a+3\rho\dot a + 3p\dot a=0.
    \label{23}
\end{equation}
Taking into account as before that $\rho \propto T^4$ yields the ODE
\begin{equation}
    \frac{\dot T}{T} +\frac{\dot a}{a}=0.
\end{equation}{}
It can be easily seen that the solution of the above differential equation is
\begin{equation}
    a(T)=\frac{C}{T},
    \label{ECg0}
\end{equation}
where C is a positive arbitrary integration constant. For very large
temperatures, the scale constant tends to zero. However, in view of the
result we obtained with the critical cosmological constant, this does not
imply that the universe is singular. Therefore, we reiterate the mantra according to which the thermodynamical
method is not always a suitable tool to probe into the nature of the
initial singularity.

\subsection{Critical cosmological constant analysis for the EC cosmology}
In this section, we analyze the EC cosmology in the presence of the cosmological constant and without the restriction of
spatial flatness. Our goal is to arrive at an expression for the effective potential
which allows to calculate a critical cosmological constant in close analogy to what we have done in the context of GR.
To this end, we consider the second Friedman equation in the EC framework in the form 
\begin{equation}
    \dot a^2 -\frac{1}{3}\kappa\bar\rho a^2-\frac{\Lambda}{3} a^2=-\frac{k}{R_0^2}.
\end{equation}
The second ingredient needed in the analysis is the parametric
   expression for the density, i.e., $\rho=\rho(a)$.
It is possible to prove that the standard result found in GR
\begin{equation}\label{rho-a}
    \rho=\rho_0a^{-3\gamma},
\end{equation}
holds also in EC. In the Friedmann equations above, we took a special
relation between the spin density and the Fermionic effective density. 
In a general case, this should be related only to the density of
particles. Therefore, we replace $n_f$ with $n$ in this
section. We are interested to know how the scale factor is related to the aforementioned density. To find such a  dependence, we consider the following relation
between the particle number density and the energy density of the fluid \cite{PoplawskiInflation,Nurgalev:1983vc,PhysRevLett.56.2873}
\begin{equation}
    \frac{dn}{n}=\frac{d\rho}{\rho+p}.
\end{equation}
Taking into account the EOS $p=(\gamma-1)\rho$, a solution for $n$ can be found integrating the above equation. More precisely, we find that
\begin{equation}\label{cici}
    n=B_w\rho^{\frac{1}{\gamma}},\quad
    B_w=\frac{\rho_0^{\frac{1}{\gamma}}}{n_0}.
\end{equation}
Making use of (\ref{cici}) in (\ref{YYYYYYY}) for 
$\gamma=\frac{4}{3}$ results in 
\begin{equation}
    3(\rho-\alpha B_w\rho^{3/2})a^2\dot a+ \dot\rho a^3-3\alpha B_w\rho^{1/2}\dot \rho a^3=0
  \end{equation}
which after some calculations can be simplified to 
\begin{equation}
    \frac{d\rho}{\rho}=-4\frac{da}{a},
\end{equation}
yielding the following solution
\begin{equation}\label{arho}
    \rho=\rho_0a^{-4}.
\end{equation}
As we know, $\bar\rho$ is an effective energy density consisting of
$\rho$ and $n$ such that replacing the parametric solution just found in the second Friedmann equation yields 
\begin{equation}
    \dot a^2-\frac{1}{3}\kappa(\rho_0a^{2-3\gamma}-\alpha n^2a^2)-\frac{\Lambda}{3}a^2=-\frac{k}{R_0^2}.
    \label{ec22}
\end{equation}
On the other hand, combining (\ref{cici}) with (\ref{arho}) leads to the following equivalent form of the second Friedmann equation, namely 
\begin{equation}
    \dot a^2-\frac{1}{3}\kappa\left(\rho_0a^{2-3\gamma}-\alpha B_w^2\rho_0^{\frac{2}{\gamma}}a^{-4}\right)-\frac{\Lambda}{3}a^2=-\frac{k}{R_0^2}.
    \label{ec2}
\end{equation}
If we introduce the effective potential 
\begin{equation}
    V_c(a)=-\frac{1}{3}\kappa\left(\rho_0a^{2-3\gamma}-\alpha B_w^2\rho_0^{\frac{2}{\gamma}}a^{-4}\right)-\frac{\Lambda}{3}a^2,
    \label{ECPPPP}
\end{equation}
then, we can rewrite (\ref{ec2}) as 
\begin{equation}
    \dot a^2+ V_c(a)=-\frac{k}{R_0^2}.
    \nonumber
\end{equation}
To find the conditions for nonsingular universes, we will analyze the shape of $V_c(a)$ for different choices of $\Lambda$. In the case of radiation, (\ref{ECPPPP}) reads
\begin{equation}
    V_c(a)=-\frac{1}{3}\kappa\left(\rho_0a^{-2}-\alpha B_w^2\rho_0^{\frac{3}{2}}a^{-4}\right)-\frac{\Lambda}{3}a^2.
    \label{ECpotg}
  \end{equation}
 We observe that for $\Lambda<0$ the potential will approach $\infty$ for $a\rightarrow 0$ as well as for $a\rightarrow \infty$. Nonsingular bound universes will then emerge (for $k<0$) for a certain $\Lambda_c$ since there is a local minimum at $a_{min}$.
\begin{figure}[H]
    \centering
    \includegraphics[scale=0.5]{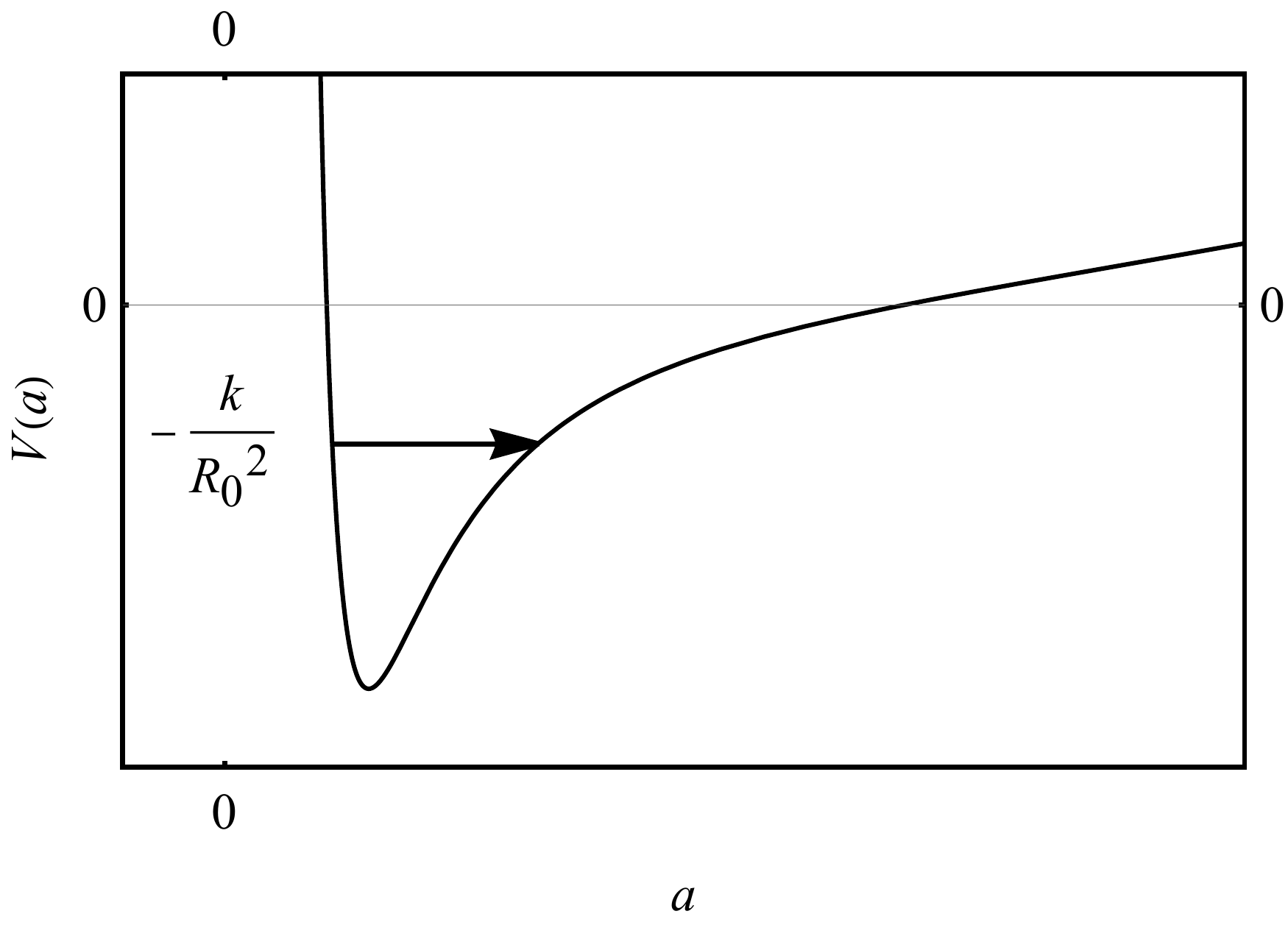} 
    \caption{The plot represents the effective potential (\ref{ECpotg}) versus $a$ in arbitrary units for $\gamma=\frac{4}{3}$ and $\Lambda<0$. $V_c$ will have a minimum value at $a_{min}$.}
    \label{EC001}
\end{figure}
Its value is determined by imposing that ${V}^{\prime}_c(a)=0$. A straightforward computation yields the following algebraic equation
\begin{equation} \label{cubic}
    Ca_{min}^6+Aa_{min}^2=B,\quad
    A=\frac{2}{3}\kappa\rho_0,\quad
    B=\frac{4}{3}\kappa\alpha B_w^2\rho_0^{3/2},\quad
    C=2\frac{|\Lambda|}{3}
\end{equation}
By means of the substitution $x=a_{min}^2$ it can be reduced to a cubic equation for $x$ . It is not necessary at this stage to search for a
full solution of the cubic equation. We will limit ourselves to the
qualitative analysis based on the shape of the potential. To this purpose, it is instructive to consider the motion reality condition 
\begin{equation}
    -\frac{k}{R_0^2}-V_c(a)\geq 0.
    \label{ECineq}
\end{equation}
As it can be evinced from Figure \ref{EC001}, (\ref{ECineq})  will never hold for $k>0$.
For $k<0$ and  $\frac{|k|}{R_0^2}-V_c(a_{min})>0$ nonsingular bound
universes will appear and their evolution is represented by the arrow. For $k<0$ and $\frac{|k|}{R_0^2}-V_c(a_{min})=0$, a stable static universe is born, which has a length $a_{min}$. Moreover, the inspection of the cubic equation (\ref{cubic})  reveals that its
discriminant $D=q^2+p^3$ with $3p=A/C$ and $2q=-B/C$ is positive. This implies that there is only one critical point corresponding to the minimum in Figure 3. Repeating the analysis for $\Lambda >0$ the discriminant can become negative which gets reflected in the two
critical points in Figure 4. In this case, the effective potential approaches $\infty$ for $a\rightarrow 0$ and $-\infty$ for $a\rightarrow \infty$.
\begin{figure}[H]
    \centering
    \includegraphics[scale=0.5]{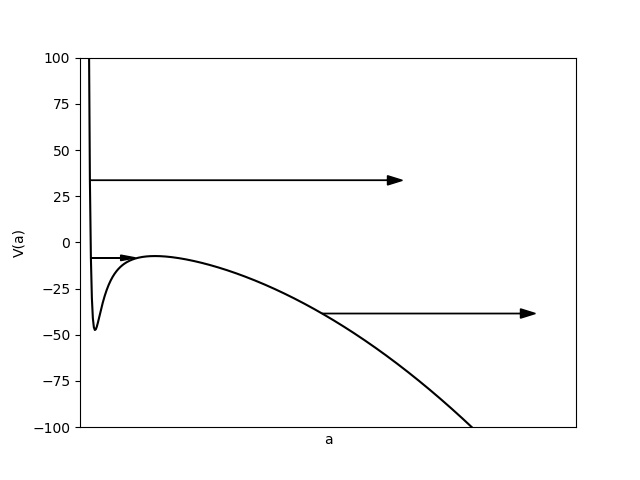} 
    \caption{The plot represents the effective potential (\ref{ECpotg}) versus $a$ in arbitrary units for $\gamma=\frac{4}{3}$ and $\Lambda>0$. As we can see, for any value of $k$, the universe represented is nonsingular.} 
    \label{EC002}
\end{figure}
For any value of $k$, (\ref{ECineq}) holds as long as $a$ is bigger than a minimum value $a_{min}$. A nonsingular universe is then born with a minimum length $a_{min}$ given by $-\frac{k}{R_0^2}=V_c(a)$. Furthermore, for $\Lambda=0$   the effective potential blows up for $a\rightarrow 0$ while it vanishes for $a\rightarrow \infty$. 
\begin{figure}[H]
    \centering
    \includegraphics[scale=0.5]{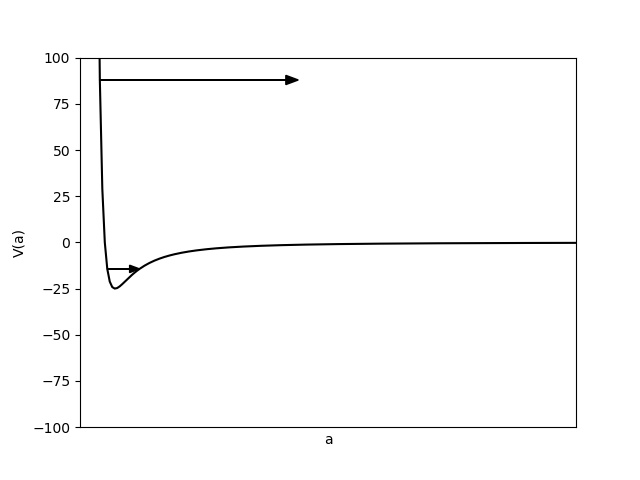} 
    \caption{The plot represents (\ref{ECpotg}) versus $a$ in arbitrary units for $\gamma=\frac{4}{3}$ and $\Lambda=0$. $V_c$ will have a minimum value at $a_{min}$.}
    \label{EC003}
\end{figure}
In addition to the previous observations, it can be easily shown that a local minimum value will appear at 
\begin{equation}
    a_{min}=\sqrt{2\alpha B_w^2\sqrt{\rho_0}}.
    \label{ECminLc}
\end{equation}
If $k<0$ for (\ref{ECineq}) to hold, the scale factor must describe a nonsingular
universe with a minimum value $a_{min}$ given by
$\frac{|k|}{R_0^2}=V_c(a)$. Its evolution is represented by the higher arrow in Figure \ref{EC003}. For $k>0$ and $-\frac{|k|}{R_0^2}-V_c(a_{min})>0$, a nonsingular bound universe will appear and its evolution is represented by the lower arrow in the same figure. Finally, for  $-\frac{|k|}{R_0^2}=V_c(a_{min})$, $a(t)$ will describe a static stable universe whose size does not change with time. In this scenario, the corresponding scale factor is given by (\ref{ECminLc}).

\section{Critical densities in EC cosmologies}
In standard flat EC cosmology as considered above, certain critical densities can be defined based on the Friedmann
equations and the continuity equation. These critical densities
suggest different behaviours in the early universe. One of which is consistent with a nonsingular universe and might also be interpreted as a bounce. In this section, we briefly explore the fact that
EC cosmology allows for three critical densities which
posit limits to the cosmological expansion and reveal particular behaviours obtained in the theory. We begin by considering equation (\ref{Spin-EC}), the parametric solution for $n$ and the corresponding results previously obtained for $\rho$. At this point, we can write the Friedman equation as given in (\ref{Friedmann1-EC}) for the case of the early universe ($\gamma=4/3$) as follows
\begin{equation}\label{schiappa}
H^{2}=\frac{\kappa}{3}\left(\rho - \frac{\kappa B_{w}}{4}\rho^{\frac{3}{2}} \right).
\end{equation}
From (\ref{schiappa}) we note that there exists a certain value for $\rho$ which makes $H=0$ and sets a maximum density since the right hand side of the equation must be positive to preserve consistency. This particular value is given by
\begin{equation}
\rho_{\rm crit}= \frac{16}{\kappa^{2}B_{w}^{2}}.
\end{equation}
To this critical density there corresponds the following value of the scale factor $a$ 
\begin{equation}
a_{\min}=\sqrt[4]{\frac{\rho_{0}}{\rho_{\rm crit}}}=\frac{1}{2}\sqrt{\kappa B_{\frac{1}{3}}\rho_{0}^{\frac{1}{2}}}
\le 1.
\end{equation}
where in the last step we used (\ref{arho}). The fact that it is possible to relate the critical density to a minimum value attained by $a$, reveals that a nonsingular bouncing scenario is admissible in the EC theory. If we further explore the other Friedmann equation,  we note that yet another 
critical density might be defined. This can be easily seen if we rewrite equation (\ref{Friedmann2-EC}) as
\begin{equation}\label{kkk}
\ddot{a}=-\frac{\kappa}{6}\left(\rho + 3p - \kappa s^{2} \right)a.
\end{equation}
With the help of (\ref{kkk}) we can explore the possibility of inflation in the early universe  \cite{PhysRevLett.56.2873,SpinInflation,PoplawskiInflation}. One possible stage of inflation also known as `super-inflation' \cite{SuperInflation,Lucchin} occurs when $\ddot{a}>0$ and $\dot{H}>0$. Using (\ref{kkk}) and considering the functional relation between $s^{2}$ and $\rho$ previously obtained for the radiation case signalizes that $\ddot{a}>0$ whenever
\begin{equation}
\rho > \frac{4}{\kappa^{2}B_{w}^{2}}\equiv \rho_{\rm inf1}.
\end{equation}
On the other hand, if we consider equations (\ref{Friedmann1-EC}) and (\ref{Friedmann2-EC}), we find that
\begin{equation}
\dot{H}=\frac{\ddot{a}}{a}-\frac{\dot{a}^{2}}{a^{2}}=-\frac{\kappa}{2}\left(\rho + p -\frac{1}{2}\kappa s^{2} \right).
\end{equation}
At this point, it can be easily seen that $\dot{H}>0$ happens whenever
\begin{equation}
\rho > \frac{64}{9 \kappa^{2}B_{w}^{2}}\equiv \rho_{\rm inf2}.
\end{equation}
Since these two densities are still below $\rho_{\rm crit}$, they cannot be ruled out by the model considered here.
As suggested by \cite{InflationGuth}, there will be an exponential stage of inflation which happens when $\dot{H}=0$, namely when the density attains the value $\rho_{\rm inf2}$. According to the results predicted by our model, we have the following stages in the early universe
\begin{itemize}
\item
if $\rho_{\rm crit}>\rho>\rho_{\rm inf2}$, we have $\ddot{a}>0$ and $\dot{H}>0$ which is referred to as the superinflation stage \cite{Lucchin}.
\item
If $\rho=\rho_{\rm inf2}$, then $\dot{H}=0$ and therefore, exponential inflation occurs.
\item
If $\rho_{\rm inf2}>\rho>\rho_{\rm inf1}$, we have a power law inflation, namely,  $a(t) \sim t^{q}$ with $q>1$  \cite{Liddle}.
\end{itemize}
The scenarios described above are peculiar to the EC cosmology \cite{PhysRevLett.56.2873, PoplawskiBounce}.  However, we would like to underline that such a model can be further generalized so that the occurrence of particular critical densities is preserved. To this purpose, let us recall that EC cosmologies can be generalized by setting up a Lagrangian with free coefficients $a_{i}$ and $b_{i}$ of contractions of independent diffeomorphic invariants. This can be achieved in two different ways leading to the Lagrangians
\begin{eqnarray}
L_{1}&=&\sqrt{-g}\left(\accentset{\circ}{R}+a_{1}K\indices{_{\rho}^{\alpha\lambda}}K\indices{_{\lambda\alpha}^{\rho}}-a_{2}K\indices{_{\alpha}^{\lambda\alpha}}K\indices{_{\rho\lambda}^{\rho}}\right),\\
L_{2}&=&\sqrt{-g}\left(\accentset{\circ}{R}+b_{1}S_{\lambda}S^{\lambda}+b_{2}S_{\alpha\beta\gamma}S^{\gamma\beta\alpha}+b_{3}S_{\alpha\beta\gamma}S^{\alpha\beta\gamma} \right)\label{torsioninv}
\end{eqnarray}
where the second Lagrangian can be also written in terms of the contorsion tensor as follows
\begin{equation}\label{contorsioninv}
L_{2}=\sqrt{-g}\left[ \accentset{\circ}{R}+\frac{b_{1}}{4}K\indices{_{\alpha\lambda}^{\alpha}}K\indices{_{\rho}^{\lambda\rho}}+\left(\frac{3b_{2}}{4}-\frac{b_{3}}{2} \right)K_{\alpha\beta\gamma}K^{\gamma\beta\alpha}
 +\left(\frac{b_{3}}{2}-\frac{b_{2}}{4} \right)K_{\alpha\beta\gamma}K^{\alpha\beta\gamma}\right].
\end{equation}
If we employ the Lagrangian $L_1$, the Friedmann equations and the continuity equation read
\begin{eqnarray}
H^{2}&=&\frac{\kappa}{3}\left(\rho+\frac{3}{4}\frac{\kappa}{a_{1}}s^{2} \right),\label{Friedmann-AEC}\\
H^{2}+\dot{H}&=&-\frac{\kappa}{6}\left(\rho+3p \right)=\frac{\ddot{a}}{a},\\
\frac{d}{dt}\left(\rho + \frac{3}{4a_{1}}\kappa B_{w}\rho^{\frac{3}{2}} \right)&=&-3H\left(\rho+p+\frac{1}{2a_{1}} \kappa B_{w}\rho^{\frac{3}{2}} \right).\label{ultima}
\end{eqnarray}
From equation (\ref{ultima}) it can be shown that
\begin{equation} \label{yyy}
a=\left(\frac{8a_{1}+3\kappa
    B_{w}\sqrt{\rho_{0}}}{8a_{1}\rho^{\frac{1}{4}}+3\kappa B_{w}
    \rho^{\frac{3}{4}}} \right)\rho_{0}^{\frac{1}{4}},
\end{equation}
that is the behaviour of the scale factor as a function of the density changed. However, since $a=a(\rho)$ and knowing $\rho=\rho(T)$, it is possible to express the scale factor in terms of the cosmological temperature. In order to extract some additional information from the first Friedmann equation, we can use (\ref{yyy}) to derive the following expression for the ratio $\dot{a}/a$
\begin{equation} \label{ccc}
\frac{\dot{a}}{a}=-\frac{2a_{1}\rho^{-\frac{3}{4}}+\frac{9}{4}\kappa B_{w} \rho^{-\frac{1}{4}}}{8a_{1}\rho^{\frac{1}{4}}+3\kappa B_{w} \rho^{\frac{3}{4}}}\frac{d\rho}{dt}
\end{equation}
which in turn allows to rewrite (\ref{Friedmann-AEC}) in such a way that a physically relevant case arises, namely
\begin{equation}
H^{2}=\frac{\kappa}{3}\left(\rho-\frac{3}{4|a_{1}|}\kappa B_{w} \rho^{\frac{3}{2}} \right).
\end{equation}
From the above expression, it is clear that a maximum critical density may arise. More precisely, we find that
\begin{equation} \label{bound2}
\rho<\frac{16a_{1}^2}{9\kappa^2 B_{w}^2}\equiv \rho_{\rm max}.
\end{equation}
If we also examine the continuity equation 
\begin{equation} \label{cont}
\dot{\rho}\left(1-\frac{9}{8|a_{1}|}\kappa B_{w}\rho^{\frac{1}{2}} \right)=-4H\rho\left(1-\frac{3}{8}\frac{\kappa B_{w}}{|a_{1}|}\rho^{\frac{1}{2}} \right),
\end{equation}
and we cast it into the form
\begin{equation}\label{CEqa1}
\dot{\rho}=-4H\rho\frac{1-\frac{\rho^{1/2}}{\rho_{c_{2}}^{1/2}}}{1-\frac{\rho^{1/2}}{\rho_{c_{1}}^{1/2}}},\quad
\rho_{c_{1}}^{1/2}\equiv \frac{8|a_{1}|}{9\kappa B_w},\quad
\rho_{c_{2}}^{1/2}\equiv \frac{8|a_{1}|}{3\kappa B_w},
\end{equation}
it becomes evident that additional critical densities may arise. It is relevant to note that the ordering of these  critical densities is
\begin{equation}
\rho_{c_{2}}>\rho_{max}>\rho_{c_{1}}.
\end{equation}
We can immediately disregard $\rho_{c_{2}}$ because it leads to an unphysical scenario, namely $H$ becomes imaginary. Moreover, $\rho_{c_{1}}$ can be disregarded as well because it would cause $\dot{\rho}$ to blow up as it can be seen from (\ref{CEqa1}). The only physically reasonable possibility is that $\rho<\rho_{c_{1}}<\rho_{\rm max}$. This avoids any contradictions but it also discards the bounce scenario. However, as it is shown in \cite{Medina:2018rnl} this does not rule out a nonsingular behaviour of the universe.

Coming back to the Lagrangian $L_2$ introduced in  (\ref{contorsioninv}), a slightly more intricate combination of the parameters $b_{i}$ appear in the Friedmann 
equations. In this case, the general form of both Friedmann equations and the continuity equation are given by the following expressions
\begin{eqnarray}
H^{2}&=&\frac{\kappa}{3}(\rho+A_{1}\kappa s^{2}),\label{FirstFriedmannALA}\\
\ddot{a}&=&-\frac{\kappa}{6}(\rho+3p-A_{3}\kappa s^{2})a,\label{SecondFriedmannALA}\\
\frac{d}{dt}\left(\rho+A_{1}\kappa B_{w}\rho^{\frac{3}{2}} \right)&=&-3H\left(\frac{4}{3}\rho+A_{2}\kappa B_{w}\rho^{\frac{3}{2}} \right).\label{ContinuityEqAL1}
\end{eqnarray}
Here, the coefficients $A_{i}$ are combinations of the coefficients $b_{i}$ and we refer to \cite{Medina:2018rnl} for the corresponding equations. As we will show, such coefficients play a role upon the classification of possible critical densities. In that regard, we can consider equation (\ref{FirstFriedmannALA}) where a maximum density emerges provided that $A_{1}<0$, i.e.
 \begin{equation}\label{rhomaxAL}
\tilde{\rho}_{\max}^{1/2}=-\frac{1}{A_{1}\kappa B_{\frac{1}{3}}}.
\end{equation}
This is the density which yields a nonsingular bounce. From the continuity equation (\ref{ContinuityEqAL1}) another critical density arises when $\rho=\frac{4}{9}\tilde{\rho}_{\rm max}$. Similarly as before, such a density leads to an inconsistency in the continuity equation and sets a new maximum on the density unless $A_{2}=2A_{1}$. Another possible critical density is represented by $\rho=\frac{16A_{1}^{2}}{9A_{2}^{2}}\tilde{\rho}_{\rm max}$ which also yields a contradiction unless this 
density is the same as the one previously mentioned. These critical densities set limits on the values of the parameters, which are constrained so that they yield the nonsingular bouncing universe.

Last but not least, an additional possibility emerges if we  examine the other Friedmann equation. More precisely, the case $\ddot{a}>0$ is of interest in the inflation scenario. In order for $\ddot{a}$ to be positive, we need to require that $\rho>\frac{4}{\kappa^2 B_{w}^2 A_{3}^2}=\tilde{\rho}_{\rm inf}$. This
density must also be smaller than the maximum density, in order for it to be consistent. In this generalized scenario,  the possibility for a family of nonsingular, bouncing and inflationary models arises. All of which are classified by the coefficients in the Lagrangian and the critical densities in the model.

\section{The Milne \& McCrea universes}
As we have seen before, the Friedmann equations are derived
introducing the FLRW metric in the Einstein field equations. However, the same feat can be done assuming the existence of expansion in the framework of Newtonian mechanics, as was firstly done by McCrea and Milne \cite{10.1093/qmath/os-5.1.73}. In our eyes, the interest in such derivation is to be linked to the possibility of including quantum corrections and to the subsequent analysis of the modified Friedmann equations. Moreover, it could well be that the inclusion of quantum corrections in the McCrea-Milne theory reveals some universal
features of the role of quantum mechanics in cosmology.
As we will see, this model recovers the same behaviour that was obtained in the EC theory of gravity for a given sign of the quantum correction. We begin with the classic derivation of the Friedmann equations, then repeat the same feat with a modified potential and at the end, we study the behaviour of the scale factor using the critical constant model and a different method. In the derivation of the Friedmann equations with and without the quantum correction, as well as in the critical density analysis, we follow \cite{Bargueno2016}.

\subsection{Friedmann equations from Newtonian dynamics}
The classical derivation starts from the consideration of an expanding universe. Such a universe can be described by Hubble's law
\begin{equation}
    \frac{dR}{dt}=HR,
    \nonumber
\end{equation}
where $R$ is the scale parameter and $H$ is the Hubble parameter. We recall that the total energy of an object of mass $m$ is described by 
\begin{equation}
    E_c= \frac{1}{2}m\left(\frac{dR}{dt}\right)^2-\frac{GMm}{R},
    \label{pot1}
\end{equation}
where $M$ can be interpreted as the total mass residing inside a sphere of radius $R$ (our universe) such that 
\begin{equation}
    M=\frac{4}{3}\pi R^3\rho.
\end{equation}
Here, $\rho$ denotes the density of the universe and $k=-2E/m$. With such considerations, (\ref{pot1}) yields the first Friedmann equation
\begin{equation}
    H^2=\frac{8\pi G}{3}\rho-\frac{k^2}{R^2}.
    \label{MMfeq1}
\end{equation}
To arrive at the second Friedmann equation, we must take into consideration a thermodynamical argument. More precisely, the  work done by the infinitesimal expansion of the universe $dV$ is $pdV$ and it decreases the energy in the volume by the same amount. Taking this into consideration and making use of the energy mass equivalence allows to conclude  that 
\begin{equation}
    R\frac{d\rho}{dt}+3(\rho+p)\frac{dR}{dt}=0.
    \label{MMcont}
\end{equation}
This is the familiar expression of the continuity equation found in the standard cosmology. It is then possible to arrive at the second Friedmann equation by considering both the continuity and the first Friedmann equation. If such procedure is followed, we arrive at the familiar result 
\begin{equation}
    \frac{d^2R}{dt^2}=-\frac{4\pi G}{3}(\rho+3p)R.
    \label{MMfeq2}
\end{equation}
This completes the derivation of the two Friedmann equations in the framework of Newtonian mechanics. It should be noted that this is not the only way to arrive at the equations and different approaches have been proposed \cite{Bargueno2016}.

\subsection{Quantum corrections to the Newtonian potential}
It has been proposed that the crucial step in this derivation is the way the Newtonian potential is used to derive the first Friedmann equation from an energy argument. Therefore, it should be consistent to take a quantum corrected potential to obtain a different set of Friedmann equations. This correction of the potential is obtained by taking gravity as an effective theory and from one-loop graviton calculations. Hence, we will now repeat the process done before by taking into account $\hbar$ corrections to the Newtonian potential (\ref{pot}). The correction is of the following form
\begin{equation}
    \phi(r)=-\frac{GM_1M_2}{r}\left(1-\gamma_q\frac{G\hbar}{r^2c^3}\right),
\end{equation}{}
where different values of $\gamma_q$ can be found in literature. For example, if the calculation is performed using only the one-particle-reducible scattering
amplitude, one obtains $\gamma_{q}=167/30\pi$ but if the full
scattering amplitude is used to define the interaction potential, one
gets $\gamma_{q}=-41/10\pi$  \cite{LivingReviews}.
We mention this explicitly since it seems that this is a decisive point on deciding the sign of $\gamma_q$ which is, in turn, important
for the behaviour of the scale factor.
We display some of the most recent values in the table below \cite{Bargueno2016}:
\begin{table}[h]
\centering
\begin{tabular}{|c|c|}
\hline
\textbf{(Year)Reference}&$\gamma_q$ \\ [3pt]\hline
(2002)\cite{Khriplovich:2002bt}&$-\frac{121}{10\pi}$ \\ [3pt]\hline
(2003)\cite{PhysRevD.67.084033}&$-\frac{41}{10\pi}$\\ [3pt]\hline
(2003)\cite{PhysRevD.68.0840055}&$-\frac{167}{30\pi}$\\ [3pt]\hline
(2007)\cite{Ross_2007}&$-\frac{41}{10}$\\ [3pt]\hline
(2007)\cite{PhysRevD.75.108501}&$\frac{107}{30\pi}$\\ [3pt]\hline
(2010)\cite{article44}&$\frac{122}{15\pi}$\\ [3pt]\hline
(2012)\cite{Donoghue:2012zc}&$-\frac{41}{10\pi}$\\ [3pt]\hline
(2015)\cite{Bjerrum-Bohr:2014zsa}&$-\frac{41}{10}$  \\ [3pt]\hline
\end{tabular}
\end{table}
\\
With this new potential, the total energy will be described by 
\begin{equation}
    E_q=E_c+\gamma_q\frac{G^2 \hbar Mm}{R^3 c^3}.
\end{equation}{}
Repeating the previous steps and introducing the Plank length $l_p=(G\hbar/c^3)^{1/2}$ leads to the first Friedmann equation with quantum corrections, namely
\begin{equation}
    H^2=\frac{\kappa}{3}\rho-\frac{\kappa}{3}\frac{l_p^2 \gamma_q}{R^2}\rho-\frac{k}{R^2}.
\end{equation}{}
In the presence of a cosmological constant, the expression above reads
\begin{equation}
    H^2=\frac{\kappa}{3}\rho-\frac{\kappa}{3}\frac{l_p^2 \gamma_q}{R^2}\rho-\frac{k}{R^2}+\frac{\Lambda}{3}.
    \label{MMqFe}
\end{equation}
Since the argument from which we deduced the continuity equation remains unchanged, the classic parametric solution for the density continues to hold. Starting with  (\ref{MMqFe}), it is possible to rewrite (\ref{MMqFe}) in terms of the density parameter $\Omega_m$ and the spatial curvature density $\Omega_{\Lambda}$, that is
\begin{equation}
    1=\left(1-\frac{l_p^2\gamma_q}{R^2}\right)\Omega_m+\Omega_\Lambda.
\end{equation}
Moreover, if we define
\begin{equation}
    \tilde\rho_{c}\coloneqq \frac{\rho_{crit}}{1-\frac{l_p^2\gamma_q}{R^2}}\approx\rho_{crit}\left(1+\frac{l_p^2\gamma_q}{R^2}\right)
\end{equation}
so that $\tilde\Omega_m=\rho/\tilde\rho_{c}$, we can cast it into the form
\begin{equation}
    1=\tilde\Omega_m+\Omega_\Lambda.
    \label{MMCR}
\end{equation}
\subsection{Critical density analysis}
In this subsection, we show that for a given value of the density there exists a turning point in the scale factor which suggests the presence of nonsingular universes. We begin by taking the parametric solution of the density for radiation $\rho=\rho_0a^{-4}$ in (\ref{MMqFe}) with $\Lambda=0$ which yields
\begin{equation}
    H^2=\frac{\kappa}{3}\rho\left(1-l_p^2\gamma_q\sqrt{\frac{\rho}{\rho_0}}\right).
\end{equation}
In the case $\gamma_q>0$, if we define a critical density of the form 
\begin{equation}
    \bar\rho_{c}=\frac{\rho_0R_0^4}{\gamma_q^2l_p^4},
\end{equation}
it can be readily seen that $H=0$. This means that there is a turning point at $\rho=\bar\rho_c$ which can be interpreted as a bounce. However, it does not imply the absence of a singularity.

\subsection{Critical cosmological constant analysis for the quantum corrected Milne \& McCrea model}
We repeat the same procedure followed in the critical constant model but applied now to the quantum corrected universes. We do so by studying the $\hbar$ corrected Friedmann equations. Let us begin by considering the general parametric equation for the density $\rho=\rho_0a^{-3\gamma}$ and let us recall that $R^2=R_0^2a^2$. Then, we can rewrite (\ref{MMqFe}) as 
\begin{equation}
    \dot a^2=\frac{\kappa}{3}\rho_0a^{(2-3\gamma)}+\frac{\Lambda}{3}a^2-\frac{\kappa}{3}\rho_0a^{-3\gamma}\frac{l_p^2 \gamma_q}{R_0^2}+\frac{k}{R_0^2}.
    \label{yaya}
\end{equation}
This time, the effective potential will then be
\begin{equation}
    V_q(a)=-\frac{\kappa}{3}\rho_0a^{(2-3\gamma)}-\frac{\Lambda}{3}a^2+\frac{\kappa}{3}\rho_0a^{-3\gamma}\frac{l_p^2 \gamma_q}{R_0^2}.
    \nonumber
\end{equation}
By introducing the effective potential
\begin{equation}
    V_q(a)=-\frac{\kappa}{3}\rho_0a^{(2-3\gamma)}\left(1-\frac{l_p^2\gamma_q}{R_0^2a^2}\right)-\frac{\Lambda}{3}a^2,
    \label{PotQ}
\end{equation}
(\ref{yaya}) takes the form
\begin{equation}
    \dot a^2+V_q(a)= -\frac{k}{R_0^2}.
\end{equation}
Since it has the same form as (\ref{classP}), we can follow the same approach used for (\ref{classP}). In particular, we analyze the asymptotic behaviour of the effective potential to search for the presence of critical values in $\Lambda$. The potential will have different shapes determined by the values of $\gamma_q$ and $\Lambda$ so that each case will be studied separately. First, if we take $\Lambda<0$ and $\gamma_q>0$, the potential will approach $\infty$ for  $a\rightarrow 0$ as well as for $a\rightarrow \infty$.
\begin{figure}[H]
    \centering
    \includegraphics[scale=0.5]{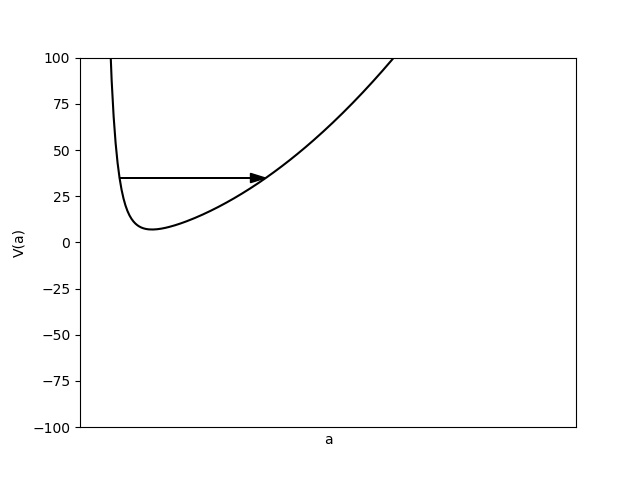} 
    \caption{The plot represents the effective potential (\ref{PotQ}) versus $a$ in arbitrary units for $\gamma=4/3$, $\Lambda<0$ and $\gamma_q>0$. $V_q$ will have a minimum value at $a_{min}$ given by (\ref{cubic1}).}
    \label{MM001}
\end{figure}
We anticipate that for $k<0$ due to the presence of an absolute minimum $a_{min}$ in the effective potential,  nonsingular bound universes will emerge (for $k<0$) in correspondence of some critical value $\Lambda_c$ of the cosmological constant. Such a minimum can be found by setting $V^{\prime}(a)=0$. This leads in the radiation case $\gamma=4/3$ to the equation
\begin{equation}
    \frac{2\kappa}{3}\rho_0a_{min}^{-4}-\gamma\kappa\rho_0\frac{l_p^2\gamma_q}{R_0^2} a_{min}^{-6}+\frac{2|\Lambda|}{3}=0.\label{puffo}
\end{equation}
If we set $x=a_{min}^{-2}$, (\ref{puffo}) becomes
\begin{equation}
    x^3-\frac{x^2}{2|B|}+\frac{2|\Lambda|}{6AB}=0,\quad
    A=\frac{2}{3}\kappa,\quad
    B=\frac{l_p^2\gamma_q}{R_0^2}.
    \label{cubic1}
\end{equation}
To find the critical value of the cosmological constant, one should first find the roots of (\ref{cubic1}) and replace them in (\ref{MMCR}) but we prefer to circumvent this cumbersome procedure in favour of a qualitative analysis. Hence, we start by imposing that the reality motion condition \begin{equation}
    -\frac{k}{R_0^2}-V_q(a)\geq 0
    \label{MMineq}
\end{equation}
must always be fulfilled. If we look at Figure~\ref{MM001}, we immediately realize that (\ref{MMineq}) never holds for $k>0$. If $k<0$ and  $\frac{|k|}{R_0^2}-V_q(a_{min})>0$, a  nonsingular bound universe may appear and its evolution is represented by the arrow in Figure~\ref{MM001}. Note that it is bound in the sense that it can attain a minimum and maximum possible value. Moreover, for $k<0$ and $\frac{-k}{R_0^2}-V_q(a_{min})=0$, a stable static universe is born which has a length $a_{min}$ given by (\ref{cubic1}). Furthermore, for $\Lambda>0$ and $\gamma_q<0$, the potential exhibits a maximum value such that the potential approaches  $-\infty$ for $a\rightarrow 0$ as well as for $a\rightarrow \infty$.
\begin{figure}[H]
    \centering
    \includegraphics[scale=0.5]{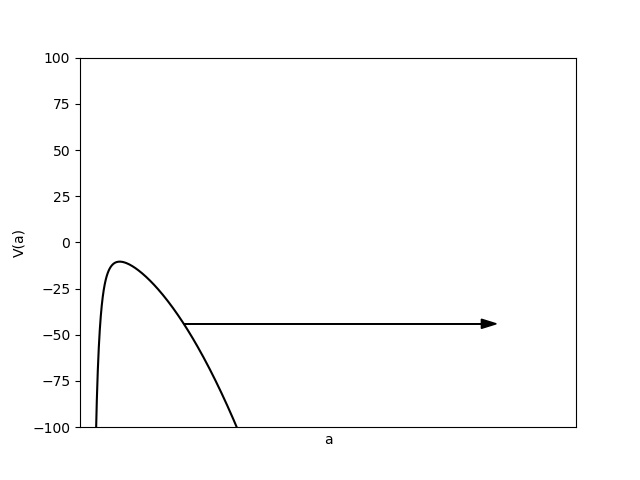} 
    \caption{The plot represents the effective potential (\ref{PotQ}) versus $a$ in arbitrary units for $\gamma=4/3$, $\Lambda>0$ and $\gamma_q<0$. $V_q$ will have a maximum value at $a$ given by (\ref{cubic2}).}
    \label{MM002}
\end{figure}
If $k>0$, nonsingular universes emerge for a certain $\Lambda_c$. In this case, following a procedure equivalent to the previous one, we obtain 
\begin{equation}
    x^3+\frac{x^2}{2|B|}+\frac{2|\Lambda|}{6A|B|}=0.
    \label{cubic2}
\end{equation}
At this point, it is instructive to take a look at Figure~ \ref{MM002} because we realize that the inequality (\ref{MMineq}) will always hold for $k<0$ such that a singular universe that expands to infinity is born. For $k>0$ and $-\frac{|k|}{R_0^2}-V_q(a_{min})>0$, depending on the place where the universe is born it can be either nonsingular expanding to infinity (collapsing and then expanding to infinity refers to the same case) or singular with a maximum length value. For $k>0$ and $-\frac{|k|}{R_0^2}-V_q(a_{min})=0$, an unstable static universe is born which has a length $a_{min}$ given by (\ref{cubic2}). For $\Lambda>0$ and $\gamma_q>0$, the potential $V_q(a)$ approaches $\infty$ for $a\rightarrow 0$ and $-\infty$ for $a\rightarrow \infty$.
\begin{figure}[H]
    \centering
    \includegraphics[scale=0.5]{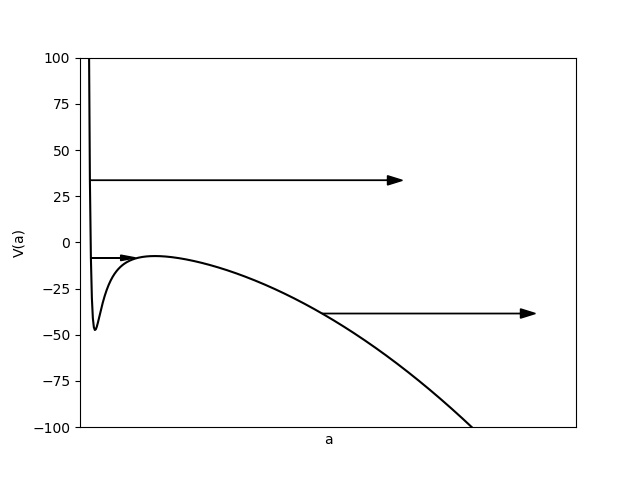} 
    \caption{The plot represents (\ref{PotQ}) versus $a$ in arbitrary units for $\gamma=4/3$, $\Lambda>0$ and $\gamma_q>0$. nonsingular universes will rise for any $k$.}
    \label{MM003}
\end{figure}
In this case, taking a look at Figure \ref{MM003}, for any value of $k$, (\ref{MMineq}) holds as long as $a(t)$ is bigger than a minimum value $a_{min}$. A nonsingular universe is then born with a minimum value given by $-\frac{k}{R_0^2}-V_q(a_{min})=0$. A static stable and bound universe is also possible for certain $k$. For $\Lambda<0$ and $\gamma_q<0$, the potential $V_q(a)$ will approach $\infty$ for $a\rightarrow \infty$ and $-\infty$ for $a\rightarrow 0$.
\begin{figure}[H]
    \centering
    \includegraphics[scale=0.5]{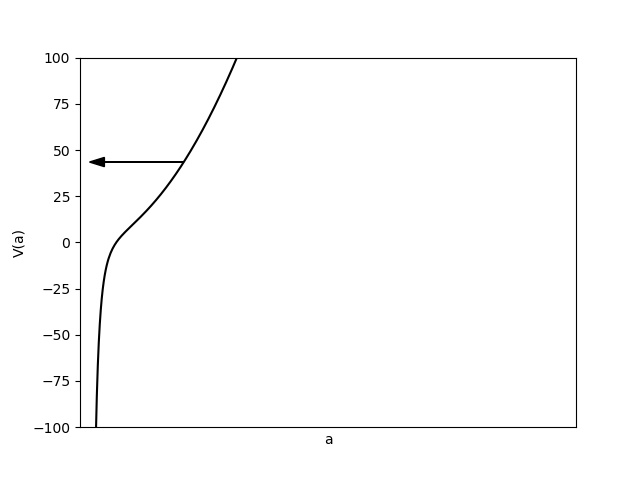} 
    \caption{The plot represents (\ref{PotQ}) versus $a$ in arbitrary units for $\gamma=4/3$, $\Lambda<0$ and $\gamma_q<0$. The universe will be singular for any $k$.}
    \label{MM004}
\end{figure}
If we inspect Figure~\ref{MM004}, we realize that  (\ref{MMineq}) holds for any value of $k$ as long as the scale factor is smaller than a maximum value $a_{max}$. A singular universe is then born with a maximum value given by $-\frac{k}{R_0^2}-V_q(a_{max})=0$. In the scenario characterized by $\Lambda=0$ and $\gamma_q>0$, the effective potential approaches $\infty$ for $a\rightarrow 0$ and vanishes for $a\rightarrow \infty$. 
\begin{figure}[H]
    \centering
    \includegraphics[scale=0.5]{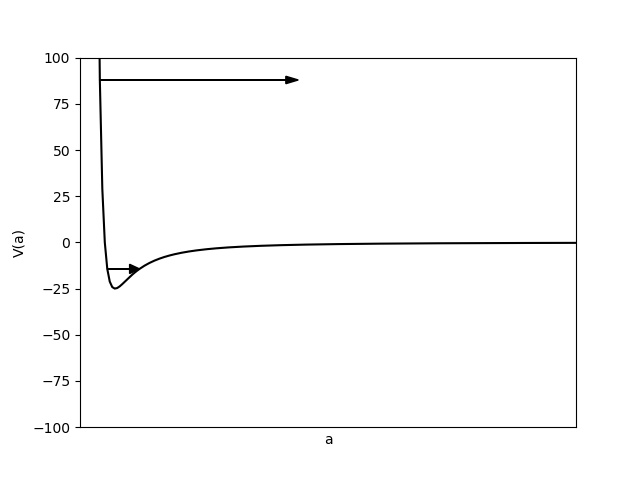} 
    \caption{The plot represents (\ref{PotQ}) versus $a$ in arbitrary units for $\gamma=4/3$, $\Lambda=0$ and $\gamma_q>0$. A stable static, bound and singular universe will be born by taking different choices of $k$.}
    \label{MM005}
\end{figure}
It can be shown that for a certain choice of $\gamma_q$, an absolute minimum value for the effective potential appears at $a_{min}$. In the radiation case, we have 
\begin{equation}
    \frac{2\kappa}{3}\rho_0a_{min}^{-4}-\frac{4\kappa}{3}\rho_0\frac{l_p^2\gamma_q}{R_0^2} a_{min}^{-6}=0
\end{equation}
with
\begin{equation}
    a_{min}={\left(\frac{2l_p^2\gamma_q}{R_0^2}\right)}^{1/2}.
    \label{MMminimum}
\end{equation}
Let $k<0$. If we consider Figure~\ref{MM005}, then  (\ref{MMineq}) holds as long as the scale factor is bigger than a minimum value $a_{min}$ given by $-\frac{k}{R_0^2}-V_q(a_{min})=0$.  If $k>0$, for (\ref{MMineq}) to hold, $a(t)$ must describe a bound universe provided that $-\frac{|k|}{R_0^2}>V_q(a_{min})$. If instead  $-\frac{|k|}{R_0^2}=V_q(a_{min})$, the scale factor describes  a static stable universe whose size does not change with time and is given by (\ref{MMminimum}). Moreover, for $\Lambda=0$ and $\gamma_q<0$, the effective potential  approaches $\infty$ for $a\rightarrow 0$ and $0$ for $a\rightarrow \infty$. 
\begin{figure}[H]
    \centering
    \includegraphics[scale=0.5]{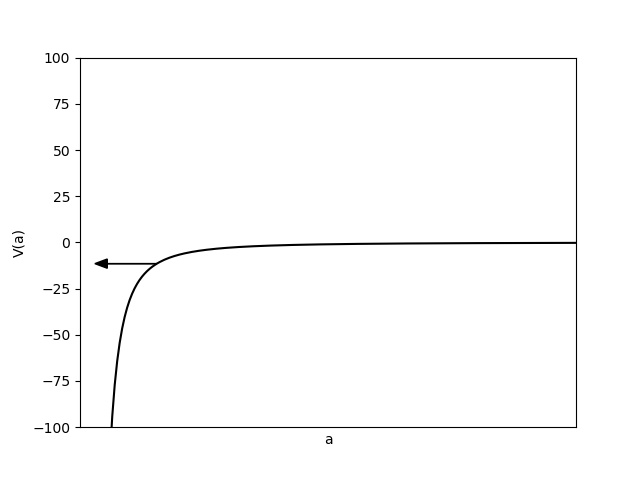} 
    \caption{The plot represents (\ref{PotQ}) versus $a$ in arbitrary
      units for $\gamma=4/3$, $\Lambda=0$ and $\gamma_q<0$.}
    \label{MM006}
\end{figure}
If we look at Figure~\ref{MM006}, we realize that (\ref{MMineq}) always holds for $k<0$. In this case, a  singular universe expanding to infinity is born. For $k>0$, (\ref{MMineq}) is satisfied as long as the scale factor is smaller than a maximum value $a_{max}$. A singular universe is then born with a maximum value obtained by solving  $-\frac{|k|}{R_0^2}-V_q(a_{max})=0$. In conclusion, we predicted the emergence of nonsingular universes depending on  certain choices for the sign of $\Lambda$ and
$\gamma_q$. Moreover, the behaviour of the potential $V_q$ highly resembles that of $V_c$ for $\gamma_q>0$. This could be due to the fact that both theories try to include quantum corrections to the classical cosmology through a geometrical feature.

\subsection{\label{sec:level1}A thermodynamical approach}
In this subsection, we repeat the thermodynamical analysis done in the EC case by finding a solution to the scale factor in terms of the temperature. To do so, we begin as usual with the two Friedmann equations with quantum corrections. If $\gamma_q>0$, they read
\begin{equation} 
    H^2=\frac{\kappa}{3}\rho-\frac{\kappa}{3}\rho\frac{l_p^2 |\gamma_q|}{R^2}-\frac{k}{R^2},\quad
    \ddot R= -\frac{\kappa}{6}(\rho+3p)R+\kappa l_p^2|\gamma_q|\frac{(\rho + p)}{2R}.
\end{equation}
Setting $R^2=R_0^2a^2$ in the above expressions gives
\begin{equation} 
    \dot a^2=\frac{\kappa}{3}\rho\left(a^2-\frac{l_p^2 |\gamma_q|}{R_0^2}\right)-\frac{k}{R_0^2},\quad
    \ddot a= -\frac{\kappa}{6}(\rho+3p)a+\kappa l_p^2|\gamma_q|\frac{(\rho + p)}{2R_0^2a}.
    \nonumber
\end{equation}
After some straightforward manipulations, reintroducing the cosmological constant and defining $A_q=l_p^2\gamma_q$ yields
\begin{equation}
    \frac{R_0^2\dot a^2+k}{R_0^2a^2}=\frac{\kappa}{3}\rho\left(1-\frac{|A_q|}{R_0^2a^2}\right)+\frac{\Lambda}{3},\quad
    \frac{\ddot a}{a}=-\frac{\kappa}{6}(\rho+3p)+\frac{\kappa |A_q|}{2}\frac{(\rho+p)}{R_0^2a^2}+\frac{\Lambda}{3}.
\end{equation}
Finally, if we multiply the first equation by $R_0^2a^2$, the second by $2R_0^2a^2$ and sum them together, we get
\begin{equation}
    2\ddot a a+\dot a^2+\frac{k}{R_0^2}=\frac{\kappa |A_q|}{R_0^2}(\frac{2}{3}\rho+p)-\kappa pa^2+\Lambda a^2. \label{N1}
\end{equation}
Now we take the first Friedmann equation with quantum corrections, multiply it by $a$ and derive it with respect to time to obtain
\begin{equation}
    \Dot a^3+2a\dot a \ddot a+\frac{k}{R_0^2}\dot a=\kappa\rho a^2\dot a+ \frac{1}{3}\kappa a^3\dot\rho +\Lambda a^2\dot a-\frac{1}{3R_0^2}\kappa\rho |A_q|\dot a.
    \label{N2}
\end{equation}{}
At this point, we multiply (\ref{N1}) by $\dot a$ to get
\begin{equation}
    2a\dot a \ddot a  +\dot a^3+\frac{k}{R_0^2}\dot a=\frac{\kappa |A_q|\dot a}{R_0^2}\left(\frac{2}{3}\rho+p\right)-\kappa pa^2\dot a+\Lambda a^2\dot a .
    \label{N3}
\end{equation}{}
and we consider the equation emerging from  (\ref{N2})-(\ref{N3}), that is 
\begin{equation}
    \left(\kappa a^2\dot a-\frac{\kappa |A_q|\dot a}{R_0^2}\right)(\rho+p)+\frac{1}{3}a^3\kappa\dot\rho=0.
\end{equation}
Using the fact that in the early universe ultrarelativistic matter is characterized by $\rho=h_eT^4$ and the EOS  $p=\rho/3$, the above equation becomes
\begin{equation}
    \left(a^2\dot a-\frac{|A_q|\dot a}{R_0^2}\right)\left(\frac{4h_eT^4}{3}\right)+\frac{4}{3}a^3(h_eT^3\dot T)=0
\end{equation}
which can be simplified to 
\begin{equation}
    \frac{\dot a}{a}-\frac{|A_q|\dot a}{R_0^2a^3}+\frac{\dot T}{T}=0.
\end{equation}
A straightforward integration leads to the solution
\begin{equation}\label{Aqpos}
    T=\frac{C}{a}\mbox{exp}\left(-\frac{ |A_q|}{2R_0^2a^2}\right)
\end{equation}
where $C$ is a positive but otherwise arbitrary integration constant. Since the function is not injective, to find the inverse of the form $a(T)$, one must restrict the domain of the function. This is done by taking two branches of the function, one from zero to the maximum value of $T(a)$, here denoted by $T_{max}$, and the other from $T_{max}$ to infinity. The first branch shows that the temperature of the universe increases with the scale factor, which signalizes an  unphysical behaviour. Hence, we focus only on the second one. If the restriction parts from the maximum value of $T(a)$ to infinity, then $a(T)$ will never be zero so a bounce is obtained.
\begin{figure}[H]
    \centering
    \includegraphics[scale=0.5]{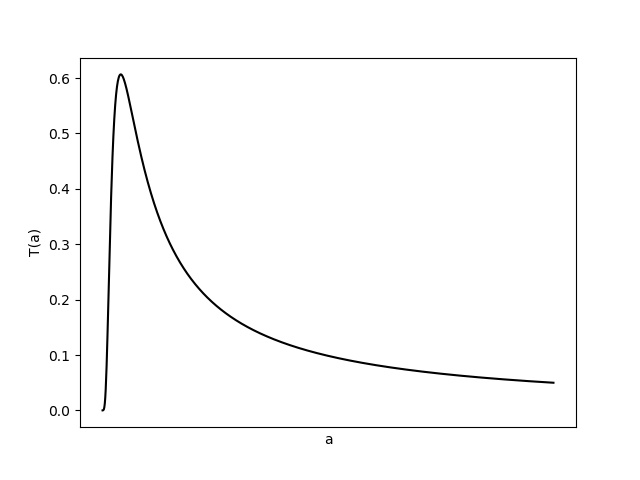} 
    \caption{Evolution of the scale parameter with respect to the temperature for $A_q$ positive in quantum corrected universes as obtained in (\ref{Aqpos}).}
    \label{fig111}
\end{figure}
Now, for $\gamma_q<0$ we have
\begin{equation}\label{Aqneg}
    T=\frac{C}{a}e^{\frac{|A_q|}{2R_0^2a^2}}.
\end{equation}{}
In this case, we have an injective function that tends to zero as $T$ decreases. It can be shown without calculations that the scale factor may describe a singular universe.
\begin{figure}[H]
    \centering
    \includegraphics[scale=0.5]{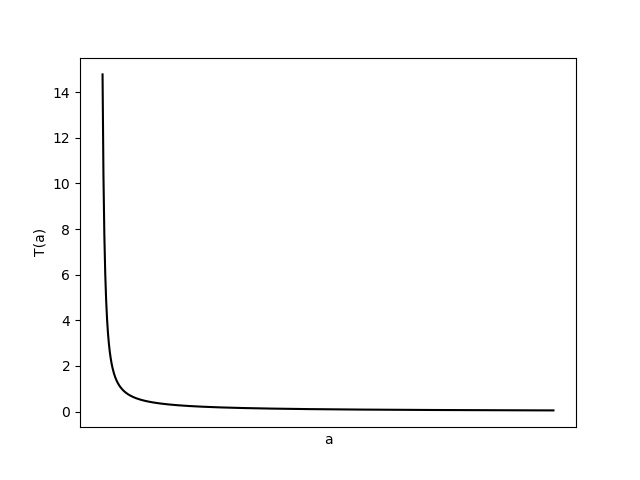} 
    \caption{Evolution of the scale parameter with respect to the temperature for $A_q<0$ in quantum corrected universes as obtained in (\ref{Aqneg}).}
    \label{fig222}
\end{figure}
Once again, we have reiterated the mantra according to which the thermodynamical method can miss crucial details of the model.

\section{Viscous Cosmologies}
Up to this point, we have treated the standard cosmology and certain models which have some formal differences with respect to the former one. EC corrections were due to modifications in the geometrical part of the Einstein Field equations but we still need to investigate how a different choice of the energy-momentum tensor may influence the cosmological model. In this section, we analyze the consequences of including bulk viscosity terms in the energy-momentum tensor and its effect in our search for nonsingular universes. In the derivation of the new Friedmann equations with bulk viscosity as well as in the identification of the new energy-momentum tensor, we follow  \cite{Bravo_Medina_2019}.

\subsection{Viscous energy-momentum tensor}
Thus far we have used the energy-momentum tensor of a perfect fluid such that the shear and bulk contributions are neglected. From now on we denote this perfect fluid tensor as $\accentset{\circ}{T}$ so that a general form of this can be written as $T_{\mu\nu}=\accentset{\circ} T_{\mu\nu}+\Delta T_{\mu\nu}$ where $\Delta T_{\mu\nu}$ is the portion including the viscous terms and is given by \cite{weinbergbook}
\begin{equation}
    \Delta T^{\alpha\beta}=-\eta h^{\alpha\mu}h^{\beta\nu}\left(\partial_\nu u_{\mu}+\partial_{\mu}u_\nu-\frac{2}{3}\eta_{\mu\nu}\partial_\sigma u^{\sigma}\right)-\zeta h^{\alpha\beta}\partial_\sigma u^\sigma
    -\chi(h^{\alpha\mu}u^\beta+h^{\beta\mu}u^\alpha)(\partial_\mu T+Tu^\sigma\partial_\sigma u_\mu).
\end{equation}
Here, $h^{\alpha\beta}=\eta^{\alpha\beta}+u^{\alpha}u^{\beta}$, $\eta^{\alpha\beta}$ is the Minkowski metric, $\eta$ is the shear viscosity coefficient, $\zeta$ is the bulk viscosity coefficient, $\xi$ is the heat conduction coefficient and $\chi$ is due to a relativistic term. In the context of general relativity, several of these expressions must be modified. The Minkowski metric must be replaced by a general metric $g_{\mu\nu}$ and partial derivatives must be replaced by covariant derivatives ensuring that if a comoving frame of reference for the fluid is considered, then the following expression for the spatial part of $\Delta T_{\mu\nu}$ holds \cite{Weinberg:1971mx}
\begin{equation}
    \Delta T^{ij}=-\eta\left(\accentset{\circ}{\nabla}^ju_i+\accentset{\circ}{\nabla}^iu_j-\frac{2}{3}\nabla_\mu u^{\mu}g_{ij}\right)-\zeta\accentset{\circ}{\nabla}_\mu u^{\mu}g_{ij}.
\end{equation}
Let us consider the FLRW metric and expand the terms in $\Delta T_{\mu\nu}$ by using  
\begin{equation}\label{uh}
    \accentset{\circ}{\nabla}_\mu u^\mu=\frac{1}{\sqrt{-g}}(\partial_\mu\sqrt{-g})u^\mu+\partial_\mu u^{\mu}.
\end{equation}
First of, if we assume an incompressible fluid, i.e.  $\partial_\mu u^{\mu}=0$ and recall that for the FLRW metric $\sqrt{-g}=r^2\sin^2(\theta) R^3(t)$, then (\ref{uh}) can be reduced to 
\begin{equation}
    \accentset{\circ}{\nabla}_\mu u^\mu=3H.
\end{equation}
Moreover, the spatial covariant derivative is
\begin{equation}
    \accentset{\circ}{\nabla}_j u_i=(\partial_j u_i-\accentset{\circ}{\Gamma}^{\lambda}_{ji}u_\lambda)=\accentset{\circ}{\Gamma}^{0}_{ji}.
\end{equation}
If we define $\bar g_{ij}=\mbox{diag}(1,r^2,r^2\sin^2{\theta})$, then the corresponding Christoffel symbols are  $\accentset{\circ}{\Gamma}^0_{ij}=R(t)\dot{R}(t)\bar g_{ij}$. Adding these terms leads the full form of the energy-momentum tensor 
\begin{equation}
    T_{ij}=(pR^2-3\zeta HR^2)\bar g_{ij}.
\end{equation}

\subsection{Friedmann equations}
Setting the energy momentum tensor into the Einstein equations yields the Friedmann equations for viscous universes. We consider the special case when $\Lambda=0$. As before, we focus our attention on the time-time and space-space components, separately. As we can see, the corrections to the perfect fluid tensor only have nonzero values on the spatial components such that the first Friedmann equation remains the same
\begin{equation}
    \left(\frac{\dot a}{a}\right)^2=\frac{\kappa}{3}\rho-\frac{k}{R_0^2a^2}.
    \label{VCCfeq1}
\end{equation}
The second Friedmann equation is obtained by taking the space-space components. We get
\begin{equation}
    -2\frac{\ddot R(t)}{R(t)}-\frac{\dot R^2(t)}{R^2(t)}=\kappa (p-3\zeta H).
\end{equation}
This is the same equation we obtained in the standard cosmology for a corrected pressure of the form 
\begin{equation}
    p'= p-3\zeta H.
\end{equation}
With the help of the relation above we can obtain the second Friedmann equation in terms of the bulk viscosity coefficient. However, we wish to find the correlation it has with the energy density $\rho$. To do so, we briefly study an equation for the bulk viscosity coefficient $\zeta$ in terms of the energy density \cite{Weinberg:1971mx,Misner:1967uu}
\begin{equation}
    \zeta=4 \bar a T^4\tau\left[\frac{1}{3}-\left(\frac{\partial p}{\partial \rho}\right)_n\right],\label{braida}
\end{equation}
where $\bar a$ is the Stefan-Boltzmann constant, $T$ is the temperature and $\tau$ is the mean free time. In view of the (\ref{braida}), it is reasonable to assume a functional relation where the bulk viscosity coefficient is proportional to the energy density, i.e. $\zeta=\alpha\rho$. This choice of $\zeta$ implies a small deviation of the EOS. Namely, if we were to take a strict EOS for radiation, $\zeta$ would yield zero. This implies that the equation which we took admits small corrections to this EOS of the form $p=(1/3+\epsilon)\rho$. Taking this into account, if we take an EOS of the form $p=(\gamma-1)\rho$ as before, then the corrected pressure will be of the form
\begin{equation}
    p'=\left(\gamma-1-3\alpha\frac{\dot a}{a}\right)\rho. 
\end{equation}
Therefore, the second Friedmann equation in terms of the dimensionless scale factor yields 
\begin{equation}
    \frac{\ddot a}{a}=-\frac{\kappa}{6}(\rho+3p').
    \label{VCCfeq2}
\end{equation}

\subsection{Critical cosmological constant analysis for the bulk viscosity model}
\begin{figure}[H]
    \centering
    \includegraphics[scale=0.5]{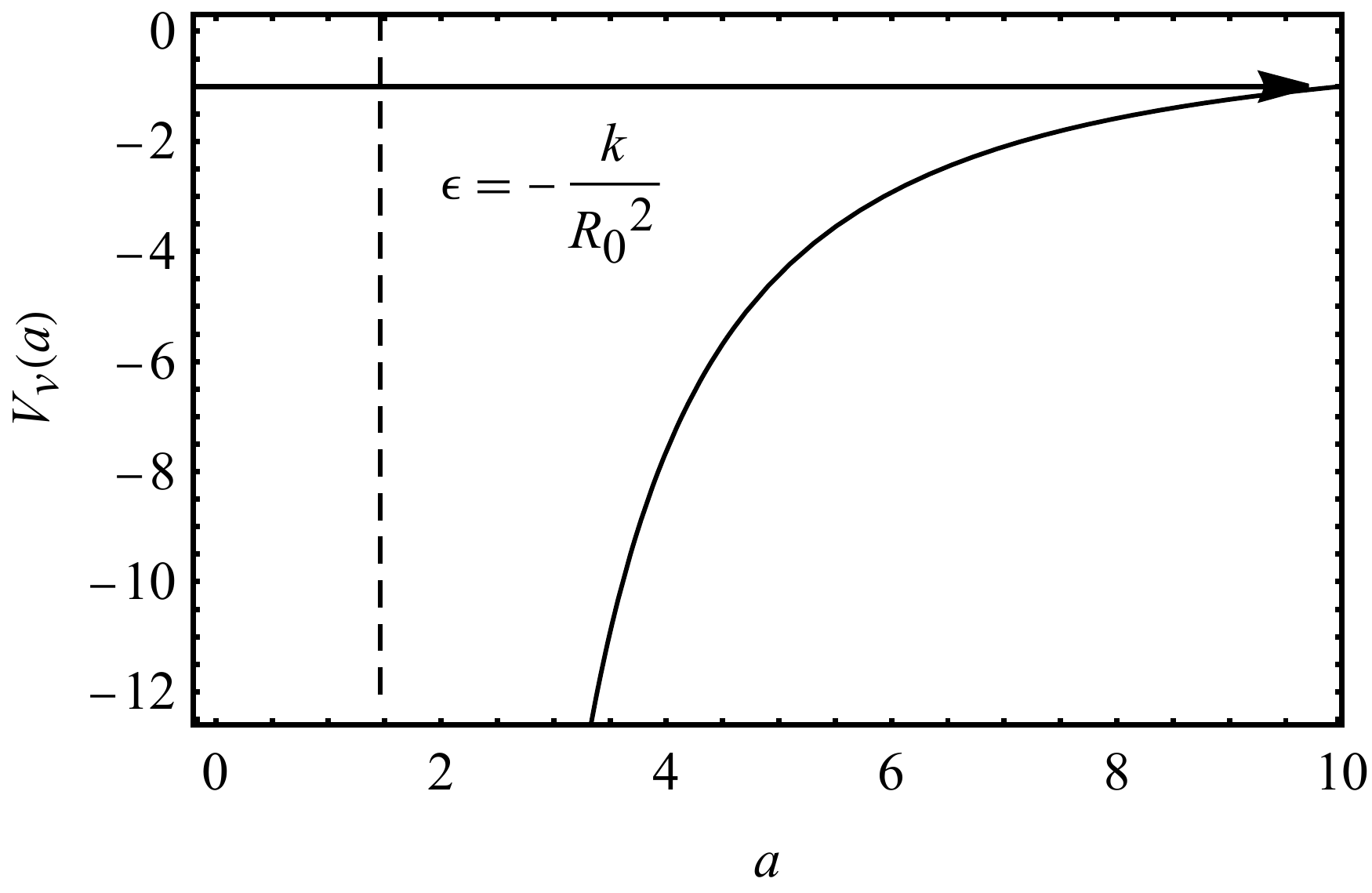} 
    \caption{Numerical solution to (\ref{VCDF}) for upper sign of the effective potential $V_v(a)$.}
    \label{UpperSignPotential}
\end{figure}

\begin{figure}[H]
    \centering
    \includegraphics[scale=0.5]{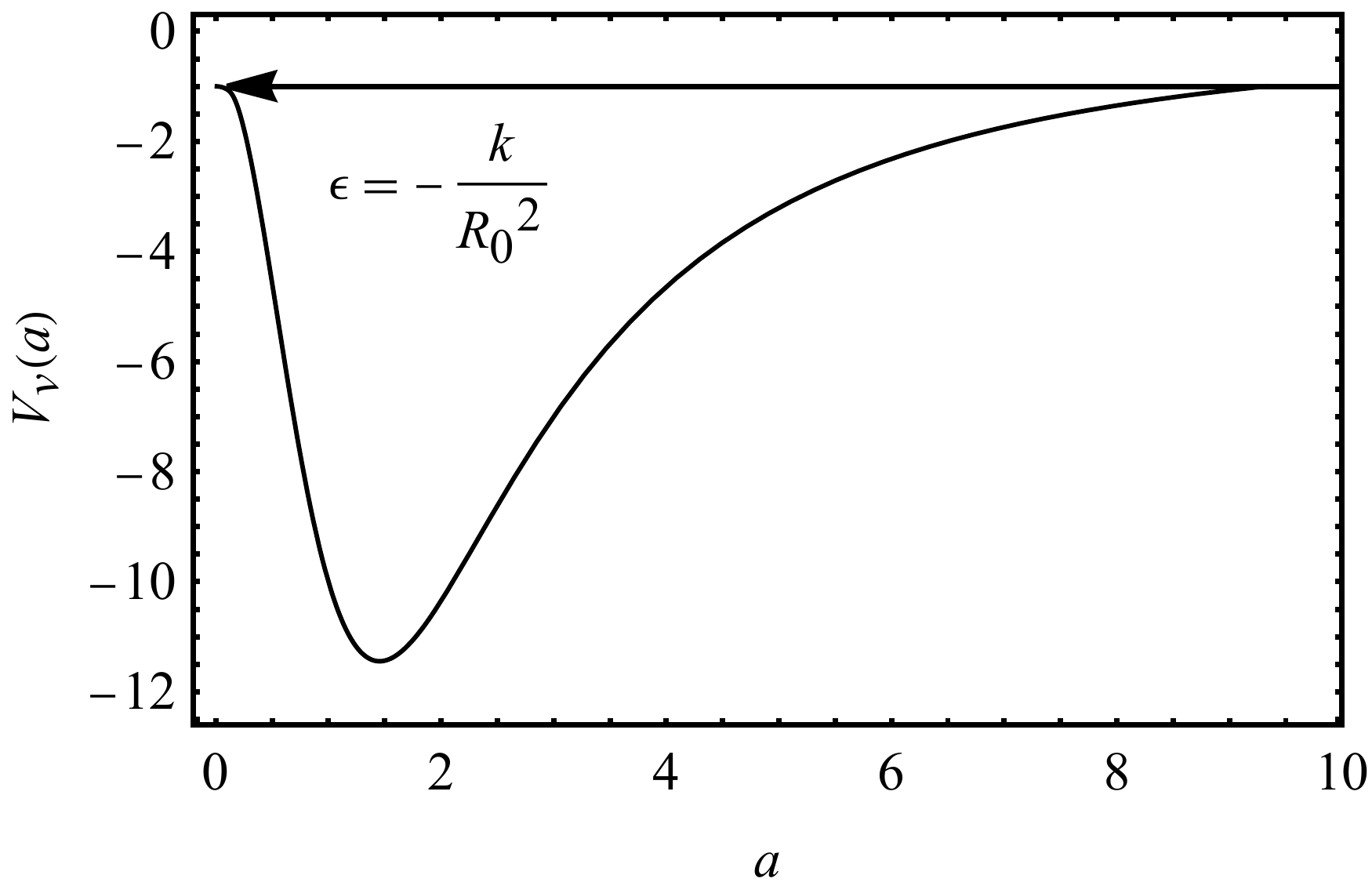} 
    \caption{Numerical solution to (\ref{VCDF}) for the lower sign of the effective potential $V_v(a)$.}
    \label{LowerSignPotential}
\end{figure}

From the  Friedmann equations(\ref{VCCfeq1}) (\ref{VCCfeq2}), we can derive the following continuity equation
\begin{equation}
    \dot \rho+3H\rho(\gamma-3\alpha H)=0.
    \label{conteqV}
\end{equation}
Also in this case, the functional dependence of $\rho$ in terms of the $a$ is in principle found by solving the equation above. However, if $k \neq 0$, solving the corresponding ODE for $\rho$ is not  trivial task. First of all, if we use the first Friedmann equation, we can rewrite (\ref{conteqV}) as 
\begin{equation}
    \frac{d\rho}{da}+3\frac{\rho}{a}\left(\gamma\pm 3\alpha\sqrt{\frac{\kappa}{3}\rho-\frac{k}{a^2R_0^2}}\right)=0,
    \label{FA}
\end{equation}{}
In the case $k=0$, the above equation simplifies to
\begin{equation}
    \frac{d\rho}{da}+3\gamma\frac{\rho}{a}\pm9\alpha\frac{\rho}{a}\sqrt{\frac{\kappa}{3}\rho}=0
\end{equation}
and it can be readily solved by separation of the variables. The corresponding solution for $\gamma=4/3$ is
\begin{equation}
    \rho^{k=0}(a)=\frac{\rho_0}{[a^2\pm\frac{9}{4}\alpha\sqrt{\frac{\kappa\rho_0}{3}}(1-a^2)]^2}.
\end{equation}
In order to study the case $k\neq0$, it is convenient to rewrite the continuity equation (\ref{conteqV}) as a differential equation involving some effective potential $V_v(a)$. This is achieved by considering the first Friedmann equation which can be cast int the form
\begin{equation}\label{bubu}
    \dot a^2 + V_v(a)= -\frac{k}{R_0^2},\quad
V_v(a)=-\frac{\kappa}{3}\rho(a)a^2.
\end{equation}
For $k=0$, the effective potential reads
\begin{equation}
    V^{k=0}_v(a)=-\frac{\kappa\rho_0}{3}\left[\frac{a}{a^2\pm\frac{9}{4}\sqrt{\frac{\kappa\rho_0}{3}}\alpha(1-a^2)}\right]^2
\end{equation}{}
 From the second equation in (\ref{bubu}) we find that
\begin{equation}\label{rhoa}
    \rho(a)=-\frac{3V_v(a)}{\kappa a^2}
\end{equation}
from which it follows that $V_{v}(a)$ is always negative. If we differentiate (\ref{rhoa}) with respect to $a$ and replace it into (\ref{FA}), we can rewrite the continuity equation as \begin{equation}
    V_v'(a)+(3\gamma-2)\frac{V_v(a)}{a} \pm \frac{9\alpha}{a^2}V_v(a)\sqrt{-V_v(a)-\frac{k}{R_0^2}}=0.
    \label{VCDF}
\end{equation}
Note that the presence of the square root in the equation above requires that we introduce the reality condition
\begin{equation}\label{real_cond}
   - \frac{k}{R_0^2}-V_v(a)\geq0.
\end{equation}
Before we study the behaviour of the solution to (\ref{VCDF}), we observe that the aforementioned equation admits the constant solution
\begin{equation}
V_v(a)=-\frac{k}{R_0^2},
\end{equation}
only in the case  $\gamma=2/3$. To study the asymptotic behaviour of (\ref{VCDF}) for $a\gg 1$, it is convenient to introduce the variable transformation $a=1/x$ mapping infinity to zero. Then, our ODE becomes
\begin{equation}\label{rapr}
\dot{V}_v(x)-\frac{3\gamma-2}{x}V_v(x)\left[1\pm x\frac{9\alpha}{3\gamma-2}\sqrt{-V_v(x)-\frac{k}{R_0^2}}\right]=0
\end{equation}
where a dot means differentiation with respect to the variable $x$. If we assume that $V_v$ is bounded as $x\to 0$, then its behaviour for $x\ll 1$ is captured by the ODE
\begin{equation}
\dot{V}_{v,0}(x)-\frac{3\gamma-2}{x}V_{v,0}(x)=0.
\end{equation}
Solving the above first order linear and homogeneous differential equation and switching back to the independent variable $a$ leads to the following  representation of the solution to (\ref{VCDF}), here denoted by the symbol $V_{v,\infty}$, valid for large values of the scale factor $a$, namely
\begin{equation}
V_{v,\infty}(a)=c_1 a^{2-3\gamma}
\end{equation}
with $c_1$ an arbitrary integration constant. In the special case $\gamma=4/3$, we have $V_{v,\infty}(a)=c_1/a^2$. In order to derive an approximated solution for small $a$, it is again convenient to work with the representation (\ref{rapr}). Since $a\ll 1$ corresponds to $x\gg 1$, we immediately see that (\ref{rapr}) reduces to the separable ODE
\begin{equation}
\dot{V}_v(x)\mp 9\alpha V_v(x)\sqrt{-V_v(x)-\frac{k}{R_0^2}}=0.
\end{equation}
In order to integrate the above equation, we introduce the auxiliary function
\begin{equation}
\psi^2(x)=-V_v(x)-\frac{k}{R_0^2},
\end{equation}
which is non-negative due to the condition (\ref{real_cond}). Hence, we obtain
\begin{equation} \label{integralpsi}
\int\frac{d\psi}{\psi^2+\frac{k}{R_0^2}}=\pm\frac{9}{2}\alpha x+c_2
\end{equation}
with $c_2$ an arbitrary integration constant. At this point, we need to consider the following cases:
\begin{enumerate}
\item
Case $k=0$: the approximated solution reads
\begin{equation}
V_v(a)=\frac{a}{\pm\frac{9}{2}\alpha+c_2 a}.
\end{equation}
\item
Case $k>0$ ($k=1$): we have
\begin{equation}\label{soluzia}
V_v(a)=-\frac{k}{R_0^2}\sec^2{\left[\frac{\sqrt{k}}{R_0}\left(\pm\frac{9\alpha}{2a}+c_2\right)\right]}.
\end{equation}
As it can be easily checked, this solution has infinitely many singularities located at
\begin{equation}
a_m=\pm\frac{9\alpha}{\pi R_0(1+2m)-2c_2\sqrt{k}},\quad m\in\mathbb{Z},
\end{equation}
and piling up at $a=0$. Furthermore, (\ref{soluzia}) exhibits infinitely many maxima at
\begin{equation}
a_{max,m}=\pm\frac{9\alpha\sqrt{k}}{2(\pi R_0 m-c_2\sqrt{k})},\quad m\in\mathbb{Z},
\end{equation}
with $a_{max,m}\in(a_{m+1},a_m)$. Note that $V_v$ is increasing on the open interval $(a_{m+1},a_{max,m})$ and decreasing on $(a_{max,m},a_m)$. We recall that the approximated solution (\ref{soluzia}) was obtained by assuming that $a\ll 1$. This means that the $a_0$ chosen to set up a certain initial value for $V_v$ must be taken so that $a_0\ll 1$. For such an $a_0$ we can always find an interval $(a_{m+1},a_m)$ by choosing $m$ large enough. If $a_0\in(a_{max,m},a_m)$, we expect 
 that the potential starts at the value $V_v(a_0)$ and decreases as we move away from $a_0$. On the other hand, if $a_0\in (a_{m+1},a_{max,m})$, then as we depart from $a_0$, the potential increases, reaching a maximum after which it decreases. 
\item
Case $k<0$ ($k=-1$): We do not get any useful information from our potential $V_{v}(a)$ since the inequality (\ref{real_cond}) is always satisfied as $V_{v}$ is always negative. In other words the universe in such a case is singular. For completeness, we briefly sketch the solution below. Since $V_{v}<0$ we obtain from the definition of $\psi^{2}$ the inequality $|\psi|>\frac{1}{R_{0}}$. From the integral (\ref{integralpsi}) we infer that
\[
\psi=\frac{1}{R_{0}^{2}}\coth \left(\mp \frac{9}{2}\alpha x +c_{3}\right),
\]
or
\[
V(a)=-\frac{1}{R_{0}^{2}}\frac{1}{\sinh^{2}\left(\mp\frac{9}{2}\alpha x+c_3\right)},
\]
for small values of $a$.
\end{enumerate}
The value of the potential at the minimum displayed in Figure~\ref{LowerSignPotential} can be directly computed from the ODE (\ref{VCDF}). If $a_e$ denotes the position of the minimum, a straightforward calculation gives
\begin{equation}
V_v(a_e)=-\frac{k}{R_0^2}-\frac{(3\gamma-2)^2 a_e^2}{81\alpha^2}.
\end{equation}
By solving the differential equation (\ref{VCDF}) numerically the form of the potential for the upper and lower sign are plotted in Figure~\ref{UpperSignPotential} and Figure~\ref{LowerSignPotential}, where a part of the behaviour of the solution is shown. We may further study the case for $k>0$ by taking the non-approximated equation (\ref{VCDF}) and testing the values for which $V_{v}'(a)=0$ to see if they correspond to a critical value. For the case  $\gamma=\frac{4}{3}$ we get two possible values for $V$
\begin{equation}
V_{v}(a_{e})=0, \qquad V_{v}(a_{e})=-\frac{4a_{e}^{2}}{81\alpha^{2}}-\frac{k}{R_{0}^{2}}.
\end{equation}
Note that the first solution must be discarded due to the fact that it does not fulfil the reality condition (\ref{real_cond}). By substituting back the second solution 
into equation (\ref{VCDF}), one notes that this is only a critical value when the lower sign is chosen and it is not difficult to verify that
\begin{equation}
V_{v}''(a_{e})=\frac{8}{81\alpha^{2}}+\frac{2k}{a_{e}^{2}R_{0}^{2}},
\end{equation}
which is always positive. Thus, $a_{e}$ corresponds to a minimum in the potential. This is evident in the plot presented in Figure~\ref{LowerSignPotential} where the numerical solution has been portrayed and indeed a minimum appears for the effective potential. In the case the upper sign is chosen, no critical value is obtained analytically. Moreover, we can further investigate the behaviour of the effective potential by considering the intersection of $\epsilon = -k/R_{0}^{2}$ with the potential $V_{v}(a)$. We do that by studying the equation $V_{v}(a_{c})=-k/R_{0}^{2}$. The latter is satisfied for \emph{both signs} whenever
\begin{equation}
V'_{v}(a_{c})=\frac{2k}{a_{c}}R_{0}^{2}.
\end{equation}
This suggests that the intersection occurs when the potential has a positive slope and thus is growing. This intersection is visible in both Figures~\ref{UpperSignPotential}  and \ref{LowerSignPotential} in the rightmost part of the plot where the $\epsilon$ arrow touches the curve of $V_{v}(a)$. However, another intersection is also evident for the lower sign plot at small values of $a$, below which the numerical solution fails. We attribute such a failure in the numerics to an impossibility to go below this value of $a$ which would correspond then to a minimum value of the scale factor. Last but not least, we note that the critical values of $a$ in the lower sign plot discussed correspond to: intersection with $\epsilon$ at both a maximum and minimum $a$, which suggests a bound solution, and $a_{e}$ which corresponds to the minimum shown in the lower sign possibility of the solution. While for the upper sign we only obtain a maximum value of $a$, we note that the smaller values of $a$ are also truncated numerically. This is represented in the plot by the dashed vertical line.
Another aspect which is worth to be discussed is how the plot of the solution changes with respect to the sign chosen in equation (\ref{VCDF}). First, we must underline that the upper
 and lower sign come from equation  (\ref{VCCfeq1})
and they may be seen in light of the behaviour of the derivative of the scale factor $\dot{a}$, meaning that the upper (initially positive) sign corresponds to a growing scale factor, namely an expanding universe. Then, it makes sense that the lower (initially negative) sign corresponds to a decreasing scale factor, i.e. a contracting universe. In light
of that, we analyze the obtained plots in a different way, observing how the expanding universe is bound at a high value of $a$ while the contracting universe is bound at both low and high values of $a$, but also keeping into account the direction of the evolution of $a$ based on the choice of sign. This is represented by the arrows in Figures \ref{UpperSignPotential}  and \ref{LowerSignPotential}. 
For the expanding universe it is clear that we obtain an upper bound from the plot and the analysis of the equations. However, our view of the behaviour of the potential is limited due to the fact that numerically the differential equation cannot be solved for small values of $a$ as it has been shown in the plot. Our attribution of this to a minimum value of $a$ unfortunately  cannot be obtained via the methods presented above. The previous
hypothesis is based on the behaviour seen in the flat universe with viscosity studied in \cite{Bravo_Medina_2019}.

\section{Conclusions}
In this paper, we have used several semi-analytical methods to probe
into different models of non-flat cosmology.  Among the models, we
examined the standard model with a critical cosmological constant
which we generalized to an EOS $p=(\gamma-1)\rho$.
A nonsingular universe emerges without invoking any quantum
corrections.
This is the importance of this model. In deriving the result we used
the potential method which allows to infer also semi-analytical
results
used later in the case of Einstein-Cartan, McCrea-Milne universe and
bulk viscosity models.  In addition to this powerful method, we
employed the thermodynamical method which allows to express the scale
factor through temperature and the critical density method. In all the
models under investigation we applied these different tools and
compared their effectiveness. It turned out that, for instance, the
thermodynamical
method albeit useful in the simplest version of the Einstein-Cartan theory,
is not always capable to reveal important details regarding the
cosmological singularity. The critical density method  requires a
careful examination, provided more critical values emerge as we
demonstrated in the generalized Einstein-Cartan cosmology.


\end{document}